\documentstyle[aps,prd,epsf,twocolumn]{revtex}

\newcommand{\be}{\begin{equation}}
\newcommand{\ee}{\end{equation}}
\newcommand{\bea}{\begin{eqnarray}}
\newcommand{\eea}{\end{eqnarray}}
\newcommand{\bdm}{\begin{displaymath}}
\newcommand{\edm}{\end{displaymath}}
\newcommand{\ul}{\underline}
\newcommand{\diff}{\mbox{$\rm{d}$}}
\newcommand{\cod}{\diff^{\dagger}}

\newcommand{\chris}[2]{\Gamma^{#1}_{\; #2}}
\newcommand{\chrishat}[2]{\hat{\Gamma}^{#1}_{\; #2}}
\newcommand{\christil}[2]{\tilde{\Gamma}^{#1}_{\; #2}}

\newcommand{\p}{\partial}
\newcommand{\Diff}{\mbox{$\rm{D}$}}

\newcommand{\we}{\wedge}
\newcommand{\sprod}[2]{\langle #1\, , \,#2 \rangle}
\newcommand{\Trace}[1]{\mbox{Tr} \left\{ #1 \right\}}
\newcommand{\idid}{1 \! \! 1}

\newcommand{\RR}{\mbox{$I \! \! R$}}

\newcommand{\bbA}{\mbox{\boldmath $A$}}
\newcommand{\bbcA}{\mbox{\boldmath ${\cal A}$}}
\newcommand{\bbF}{\mbox{\boldmath $F$}}
\newcommand{\bbG}{\mbox{\boldmath $G$}}
\newcommand{\bbK}{\mbox{\boldmath $K$}}
\newcommand{\bbL}{\mbox{\boldmath $L$}}
\newcommand{\bbP}{\mbox{\boldmath $P$}}
\newcommand{\bbS}{\mbox{\boldmath $S$}}
\newcommand{\bbT}{\mbox{\boldmath $T$}}
\newcommand{\bbV}{\mbox{\boldmath $V$}}
\newcommand{\bbQ}{\mbox{\boldmath $Q$}}

\newcommand{\taur}{\tau_{r}}
\newcommand{\tautheta}{\tau_{\vartheta}}
\newcommand{\tauphi}{\tau_{\varphi}}
\newcommand{\hatast}{\hat{\ast}}

\newcommand{\gtens}{\mbox{\boldmath $g$}}
\newcommand{\gfourtens}{\gtens}

\newcommand{\gtiltens}{\tilde{\gtens}}
\newcommand{\gtil}{\tilde{g}}
\newcommand{\nabtil}{\tilde{\nabla}}
\newcommand{\tilast}{\tilde{\ast}}
\newcommand{\laptil}{\tilde{\Delta}}
\newcommand{\nabhat}{\hat{\nabla}}
\newcommand{\laphat}{\hat{\Delta}}

\newcommand{\ghattens}{\hat{\gtens}}
\newcommand{\ghat}{\hat{g}}

\newcommand{\fourfold}{(M,\gfourtens)}

\begin{document}
\twocolumn[\hsize\textwidth\columnwidth\hsize\csname @twocolumnfalse\endcsname

\title{Perturbation theory for self-gravitating gauge fields 
       I: The odd-parity sector }

\author{O. Sarbach$^{\star}$, M. Heusler$^{\star}$, 
        and O. Brodbeck$^{\dagger}$}
\address{$^{\star}$Institute for Theoretical Physics, University of Zurich,
         CH--8057 Zurich, Switzerland\\
         $^{\dagger}$Max-Planck-Institute for Physics, Werner Heisenberg
         Institute, D--80805 Munich, Germany}
\date{\today}

\maketitle

\begin{abstract}
A gauge- and coordinate-invariant perturbation theory for 
self-gravitating non-Abelian gauge fields is developed and 
used to analyze local uniqueness and linear stability properties 
of non-Abelian equilibrium configurations.
It is shown that all admissible stationary odd-parity excitations 
of the static and spherically symmetric Einstein-Yang-Mills
soliton and black hole solutions have total angular momentum number
$\ell = 1$, and are characterized by non-vanishing
asymptotic flux integrals.
Local uniqueness results with respect to non-Abelian perturbations
are also established for the Schwarzschild and the Reissner-Nordstr\"om
solutions, which, in addition, are shown to be
linearly stable under dynamical Einstein-Yang-Mills perturbations.
Finally, unstable modes with $\ell = 1$ are also excluded for the
static and spherically symmetric non-Abelian 
solitons and black holes.
(PACS numbers: 04.25.Nx, 04.40.-b, 04.70.Bw)\\

\end{abstract}
]

\section{Introduction}
\label{section-IN}

Self-gravitating non-Abelian gauge fields admit a
rich spectrum of equilibrium configurations, which is
a consequence of the balance between the gravitational
attraction and the repulsive nature of the Yang-Mills
interaction. 
In particular, the static and spherically symmetric 
non-Abelian soliton \cite{Bartnik88} 
and black hole solutions \cite{VolkGal89} owe their existence to the 
nonlinearities of {\it both\/} general relativity and Yang-Mills 
theory. 

On the other hand, the key to the black hole uniqueness 
theorems \cite{Heusler} lies in the 
$\sigma$-model structure of the Einstein(-Maxwell) 
equations in the presence of a Killing field \cite{Mazur}, 
\cite{Neugebauer}. 
As this property ceases to exist for self-gravitating {\it non-Abelian\/} 
gauge fields \cite{MH-Livrev}, the classification of all stationary 
Einstein-Yang-Mills (EYM) soliton and black hole solutions is
necessarily a very difficult task. In particular,
the set of global charges (asymptotic flux integrals) does no longer 
uniquely characterize all possible non-Abelian
equilibrium configurations.

Induced by the work of Bartnik and McKinnon (BK)
on non-Abelian solitons \cite{Bartnik88}, various new
self-gravitating 
equilibrium configurations have been found during the last decade.
Besides the abovementioned static and spherically symmetric black holes with 
Yang-Mills hair (i.e., with vanishing Yang-Mills charges but
different metric structure than the Schwarzschild solution) 
\cite{VolkGal89}, these include soliton and 
black hole solutions in Skyrme, Higgs, dilaton and other non-linear 
field theories coupled to gravity (see \cite{VolkGal99}
for a review and references). 

Moreover, numerical \cite{KK97} and analytical \cite{RW95-1} 
studies have revealed that non-Abelian static black holes are 
not necessarily spherically symmetric -- in fact, they need 
not even be axisymmetric \cite{RW95-2}.
In addition, the non-linear nature of the Yang-Mills interaction
enables the existence of stationary, non-static black holes with 
vanishing Komar angular momentum \cite{BHSV}.
Also, the usual Lewis-Papapetrou form of the metric does not
necessarily describe all stationary and 
axisymmetric EYM black holes, that is, the {\it circularity\/} 
theorem does not generalize to space-times containing non-Abelian 
gauge fields \cite{MH-BadHf}.

The above comments suggest that it is not (yet) feasible to
completely classify the soliton and black hole  
solutions of the stationary EYM equations.
In this article we pursue, therefore, a more modest aim. That is,
we compute the complete spectrum of 
{\it stationary EYM perturbations\/} of the BK solitons 
and the corresponding black holes with hair. We do so by 
systematically developing the perturbation theory for
self-gravitating non-Abelian gauge fields. 
Following the tradition, we start with the odd-parity sector, and
defer the investigation of even-parity perturbations to a forthcoming 
publication \cite{MHOS-even}.

The gauge- and coordinate-invariant equations derived in this paper 
describe perturbations of arbitrary spherically 
symmetric EYM configurations, where the stationary and the 
dynamical sector can be treated separately if the background 
is static. In order to classify the equilibrium solutions close
to the BK solitons and the corresponding black holes, it
is sufficient to consider {\it stationary\/} excitations.
As we shall see, these are naturally analyzed in terms of 
invariant {\it metric\/} and Yang-Mills amplitudes. 

The main results of this paper concern two local uniqueness 
theorems, applying to the BK solitons and the corresponding 
black holes with hair, respectively:
We prove that all stationary odd-parity excitations of these static and 
spherically symmetric configurations are parametrized in
terms of infinitesimal asymptotic flux integrals. More
precisely, we show that the soliton and black hole
excitations found in \cite{BHSV} are the only stationary,
asymptotically flat perturbations of the
BK solitons and the corresponding black holes with hair. 
In particular, there exist no admissible regular or black
hole perturbations with total angular momentum number $\ell > 1$,
while for $\ell = 1$, the unique soliton and black hole excitations 
are those with infinitesimal electric charge and/or
infinitesimal Komar angular momentum \cite{BHSV}.
On the perturbative level, the situation is, therefore,
similar to the Abelian case, where the only admissible
stationary excitations of the Schwarzschild metric
are the Kerr-Newman modes. The above results also
establish a local version of the
circularity theorem in the odd-parity sector.

In addition to the classification of neighboring 
equilibrium configurations, we also discuss some {\it stability\/} 
issues, which require the analysis
of {\it dynamical\/} perturbations. Unfortunately, the 
gauge-invariant metric perturbations used in this paper are, 
in general, not suited to apply spectral analysis, since their evolution
is not governed by a standard pulsation operator. In a recent 
work \cite{OddLetter} we have demonstrated how to 
overcome this problem by using {\it curvature-based\/} quantities. 
A rigorous discussion of dynamical perturbations within the
{\it metric\/} approach is nevertheless possible for
some distinguished cases. These include $\ell = 1$
EYM perturbations of arbitrary background configuratiuons, 
and arbitrary EYM perturbations of embedded
Abelian configurations.

Hence, further results derived in this paper concern the 
non-Abelian stability (and local uniqueness) of 
the Schwarzschild and the Reissner-Nordstr\"om (RN) black holes,
as well as the stability properties of non-Abelian configurations
with respect to $\ell = 1$ perturbations.
In particular, we show that both the Schwarzschild and the
RN metric are linearly stable with respect to 
dynamical non-Abelian perturbations and admit no stationary 
excitations other than the (embedded) Kerr-Newman modes. 
In addition, we establish the absence of unstable modes
of the pulsation equations governing the $\ell = 1$ perturbations
of the BK solitons and the corresponding black holes with hair.
In this context it is worthwhile recalling that unstable
Yang-Mills modes with odd-parity do exist 
for $\ell = 0$ \cite{Volkov}.

The paper is organized as follows: In Section \ref{section-GE}
we briefly review the gauge-invariant approach
to odd-parity gravitational perturbations and give a 
coordinate-invariant derivation of the Regge-Wheeler (RW) equation.
In Section \ref{section-YME} we present the harmonic decomposition
of Yang-Mills fields, using a convenient method to parametrize
su(2)-valued one-forms in terms of isospin harmonics.
Taking advantage of some powerful tools developed in 
Appendix \ref{App-D}, the linearized field equations governing
arbitrary odd-parity perturbations of spherically symmetric EYM
configurations are derived in Section \ref{section-PEQ}.

As first applications, we establish the linear stability and the local
uniqueness properties of the Schwarzschild and the RN
solutions with respect to non-Abelian perturbations
in Sections \ref{section-RN} and \ref{section-Schw}, respectively.
The local uniqueness theorems for the BK solitons and the
corresponding black holes are proven in Section \ref{section-SPEYM}. 
Eventually, in Section \ref{sect-nonstationary}, 
we establish the dynamical stability of these solutions
with respect to non-spherical perturbations with $\ell = 1$. 

A variety of technical issues, such as the expressions for the
linearized Ricci tensor, the integral argument 
excluding admissible solutions of certain RW type equations,
some asymptotic expansions, the introduction of isospin harmonics, and
the construction of gauge- and coordinate-invariant Yang-Mills 
amplitudes are discussed in Appendixes \ref{App-A}-\ref{App-F}.

\section{Gravitational perturbations}
\label{section-GE}

In this section we briefly review the gauge-invariant approach 
to odd-parity gravitational perturbations \cite{GS}.
As an application we derive a coordinate-invariant version 
of the RW equation \cite{RW57}. We finally recall the
arguments establishing the stability of the Schwarzschild metric 
with respect to vacuum perturbations.

\subsection{Background expressions}
\label{subsection-GE-BE}

We are analyzing odd-parity perturbations of {\it spherically 
symmetric background\/} configurations. A spherically symmetric 
spacetime $\fourfold$ is a warped product of 
$\tilde{M} \equiv M / \mbox{SO(3)}$ and $S^{2}$ with metric 
\be
\gfourtens = \gtiltens + R^{2} \, \ghattens \, .
\label{GE-1a}
\ee
Here $\ghattens$ is the standard metric on $S^{2}$, and
$\gtiltens$ and $R$ denote the metric tensor and a 
real-valued function, respectively, defined on the 
two-dimensional pseudo-Riemannian orbit space $\tilde{M}$ 
with coordinates $x^{a}$, say. Here and in the following 
lower-case Latin indices $(a=0,1)$ refer to coordinates on 
$(\tilde{M}, \gtiltens)$, while capital Latin indices 
$(A=2,3)$ refer to the coordinates $\vartheta$ and 
$\varphi$ on $(S^{2}, \ghattens)$. The dimensional
reduction of the Einstein tensor yields 
\bea
G_{ab} & = & \frac{1}{R^{2}} \left(
2 R \laptil R + \sprod{\diff R}{\diff R} - 1 \right)
\gtil_{ab} - \frac{2}{R} \nabtil_{a} \nabtil_{b} R \, ,
\nonumber \\
G_{AB} & = & \frac{1}{2} \left( 2 R \laptil R -  
R^{2} \tilde{R} \right) \ghat_{AB} \, ,
\label{GE-1b}
\eea
where the off-diagonal components vanish, $G_{Ab} = 0$. 
The operators with a tilde and the inner product 
$\sprod{\,}{\,}$ refer to the two-dimensional 
pseudo-Riemannian metric $\gtiltens$, and $\tilde{R}$ 
denotes the Ricci scalar of $\gtiltens$.

\subsection{Coordinate-invariant amplitudes}
\label{subsection-GE-CI1}

Arbitrary perturbations of spherically symmetric background 
fields can be expanded in terms of spherical tensor harmonics. 
For odd-parity perturbations the transverse spherical vector 
harmonics, 
$S_{A} \equiv (\hatast\diff Y)_{A}$ 
form a basis of vector fields on $S^{2}$, while the harmonics
$\nabhat_{\{A} S_{B\}} \equiv \frac{1}{2}
(\nabhat_{A} S_{B} + \nabhat_{B} S_{A}$) are a basis of 
symmetric tensor fields on $S^{2}$; see Appendix \ref{App-D} for details.
(Here $\hat{\ast}$ denotes the Hodge dual with respect to the 
metric $\ghattens$, and the $Y^{\ell m}$ are the scalar spherical
harmonics, where the angular numbers $\ell$ and $m$ are suppressed 
throughout, i.e., $Y \equiv Y^{\ell m}$, 
$S_{A} \equiv S_{A}^{\ell m}$.) The odd-parity perturbations of 
$g_{\mu \nu}$ are, therefore, parametrized in terms of a scalar field
$\kappa$ and a one-form $h = h_{a} \diff x^{a}$,
\be
\delta g_{ab} = 0 , \; \; \; 
\delta g_{Ab} = h_{b} S_{A} , \; \; \; 
\delta g_{AB} = 2 \kappa \nabhat_{\{A} S_{B\}} ,
\label{GE-2}
\ee
where $\kappa$ and $h_{a}$ depend on the coordinates $x^{b}$ only.

A vector field $X = X^{\mu}\p_{\mu}$ generating an infinitesimal 
coordinate transformation with odd parity is determined by a 
function $f(x^{b})$, where
\be
X^{a} = 0 , \; \; \;  
X^{A} = f \, S^{A} = \frac{f}{R^{2}} \, \ghat^{AB} S_{B} .
\label{GE-3}
\ee
Under coordinate transformations induced by $X$ the perturbations 
of a tensor field transform with the Lie derivative of the 
corresponding background quantity with respect to $X$:
$\delta t_{\mu \nu} \rightarrow \delta t_{\mu \nu}$ $+$
${\cal L}_{X} t_{\mu \nu}$. Using
${\cal L}_{X} g_{Ab} = S_{A} R^{2} \nabtil_{b} (R^{-2}f)$ and
${\cal L}_{X} g_{AB}= 2 f \, \nabhat_{\{A} S_{B\}}$, 
the metric perturbations transform according to
\be
\kappa \rightarrow \kappa + f , \; \; \; 
\frac{h_{b}}{R^2} \rightarrow  
\frac{h_{b}}{R^2} + \nabtil_{b} 
\left( \frac{f}{R^{2}} \right) .
\label{GE-4}
\ee
In a similar way one obtains the transformation laws for the 
perturbations of the Einstein tensor. Also using the background
properties $G_{Ab} = 0$ and 
$2 G^{A}_{\; B} = G^{D}_{\; D} \delta^{A}_{\; B}$ one finds
\bea
\delta G_{Ab} & \rightarrow & \delta G_{Ab} +
G^{B}_{\; A} S_{B} \,
R^{2} \nabtil_{b} \left( \frac{f}{R^{2}} \right) ,
\label{GE-5}\\
\delta G_{AB} & \rightarrow & \delta G_{AB} +
G^{D}_{\; D} \nabhat_{\{A} S_{B\}} f \, .
\label{GE-5b}
\eea

One may now use the transformation laws for $\kappa$ and $h_{b}$ 
to construct the following coordinate-invariant components:
\be
\delta G_{ab}^{inv} \equiv \delta G_{ab} \, , \;\;\;
\delta G_{Ab}^{inv} \equiv
\delta G_{Ab} - h_{b} \, G^{B}_{\; A} S_{B} \, , 
\label{GE-6a}
\ee
and, for $\ell \neq 1$,
\be
\delta G_{AB}^{inv} \equiv
\delta G_{AB} -  \kappa \, G^{D}_{\; D} 
\nabhat_{\{A} S_{B\}} \, .
\label{GE-6b}
\ee
We recall that the scalar amplitude $\kappa$ defined in 
Eq. (\ref{GE-2}) is not present for $\ell = 1$, since then
$\nabhat_{\{A} S_{B\}}$ vanishes.
However, by virtue of Eq. (\ref{GE-5b}), this also 
implies that $\delta G_{AB}$ is
already coordinate-invariant. (In fact, $\delta G_{AB}$ vanishes 
identically for $\ell = 1$, as will be shown below.) 
Hence, for $\ell = 1$ one needs only the invariant components
defined in Eqs. (\ref{GE-6a}), which do not involve the 
amplitude $\kappa$.

As the $\delta G_{\mu \nu}^{inv}$ are invariant under 
coordinate transformations generated by $X$, the expressions 
(\ref{GE-6a}) and (\ref{GE-6b}) will only involve 
coordinate-invariant combinations of the one-form $h$ and the scalar 
$\kappa$. In fact, for $\ell \neq 1$,  $\delta G_{\mu \nu}^{inv}$ 
can be expressed in terms of the manifestly coordinate-invariant 
one-form $H$, defined by
\be
H \equiv h - R^{2} \diff \left( 
\frac{\kappa}{R^{2}} \right).
\label{GE-7}
\ee
This definition is again limited to $\ell \neq 1$.
For $\ell = 1$, where $\kappa$ is absent, we will see that
the remaining perturbation $h$ enters $\delta G_{\mu \nu}^{inv}$ 
via the invariant two-form $\diff (R^{-2} h)$ only.

\subsection{Coordinate-invariant Einstein tensor}
\label{subsection-GE-CI2}

The computation of the coordinate-invariant components
$\delta G_{\mu \nu}^{inv}$ is considerably simplified by the 
following observation: In the gauge where the scalar amplitude 
$\kappa$ vanishes, henceforth called the off-diagonal gauge (ODG),
the perturbation $h$ coincides with the coordinate-invariant 
perturbation $H$ defined in Eq. (\ref{GE-7}). (It is obvious 
from Eq. (\ref{GE-4}) that the ODG always exists and fixes 
the gauge function $f$ uniquely.) Hence, for $\ell > 1$, the 
correct invariant tensors are obtained by computing 
$\delta G_{\mu \nu}^{inv}$ in the ODG, and by substituting 
$H$ for $h$ in the resulting expressions. For $\ell = 1$ all 
perturbations are off-diagonal anyway, and one obtains the 
correct expressions in terms of the invariant quantity 
$\diff (R^{-2} h)$.

It is a straightforward task to compute $\delta G_{\mu \nu}$ 
in the ODG. Using the formulas (\ref{del-GAb}), (\ref{del-GAB}) 
and (\ref{del-Gab}) derived in Appendix \ref{App-A}, 
Eqs. (\ref{GE-6a}) and (\ref{GE-6b}) yield the expressions
\bea
\delta G_{Ab}^{inv}\mid_{ODG} &=& \frac{S_{A}}{R^{2}} 
\left\{
\nabtil^{a} \left[R^{4} \nabtil_{[b} \left(
h_{a]} R^{-2} \right) \right] + 
\frac{\lambda}{2} h_{b}
\right\} ,
\nonumber\\
\delta G_{ab}^{inv}\mid_{ODG} &=& 0 , \; \; \; 
\delta G_{AB}^{inv}\mid_{ODG} = \nabhat_{\{A} S_{B\}} \, 
\nabtil^{b} h_{b} ,
\label{GE-ODG}
\eea
where
\bdm
\lambda \equiv (\ell-1)(\ell+2).
\edm
Here we have used the background property
$2 G^{A}_{\; B} = G^{D}_{\; D} \delta^{A}_{\; B}$ and the fact 
that $\delta G_{AB}^{inv} = \delta G_{AB}$ in the ODG. Since 
$h_{b}$ coincides with the invariant amplitude $H_{b}$ in the ODG, 
we may replace $h_{b}$ by $H_{b}$ in the above expressions, which 
makes them manifestly coordinate-invariant for $\ell > 1$. 
For $\ell = 1$ the second term in the expression for 
$\delta G_{Ab}^{inv}$ vanishes, and $h_{b}$ appears only via the 
coordinate-invariant expression $\nabtil_{[b} (h_{a]} R^{-2})$. 
We therefore end up with the manifestly coordinate-invariant 
expressions
\be
\delta G_{ab}^{inv} = 0 , \; \; \; 
\delta G_{AB}^{inv} = - \cod H \, \nabhat_{\{A} S_{B\}} 
\label{GE-9b}
\ee
and
\be
\delta G_{Ab}^{inv} \diff x^{b} = \frac{S_{A}}{2 R^{2}} \left\{ \cod
  \left[ R^{4} \diff \left(R^{-2} H \right) \right] + \lambda H \right\},
\label{GE-9}
\ee
which are valid for all values of $\ell$, provided that $H$ is defined 
according to Eq. (\ref{GE-7}) for $\ell > 1$, and according
to $H \equiv h$ for $\ell = 1$. Here 
$\cod \equiv \tilast \diff \tilast$ denotes the co-differential 
operator for $p$-forms on $(\tilde{M},\gtil)$, e.g.,
$\cod H = - \nabtil^{a} H_{a}$,
$(\cod \diff H)_{b} = 2 \nabtil^{a} \nabtil_{[b} H_{a]}$.

The linearized Bianchi identity implies that
the Einstein equation for $\delta G_{AB}^{inv}$ is
a consequence of the equation for $\delta G_{Ab}^{inv}$.
In fact, the first equation is the integrability condition for 
the second one, as is obvious for vacuum perturbations: 
Applying the co-differential to
$R^2\delta G_{Ab}^{inv} = 0$ yields $\cod H = 0$, that is,
$\delta G_{AB}^{inv} = 0$. (For $\ell = 1$ this integrability
condition is void, in agreement with the fact that 
$\delta G_{AB}^{inv}$ vanishes identically.)

\subsection{Local uniqueness and
linear stability of the Schwarzschild metric}
\label{subsection-GE-LU}

As an application we consider vacuum perturbations of the Schwarzschild
metric. The relevant equation for the odd-parity sector was first derived by
Regge and Wheeler \cite{RW57}, and brought in a gauge-invariant form by
Gerlach and Sengupta \cite{GS}. A gauge-invariant approach which is based on
the Hamiltonian formalism was given by Moncrief \cite{VM}.

The {\it linear stability\/} of the Schwarzschild metric follows from
the dynamical behavior of vacuum fluctuations. In order to
establish the {\it local uniqueness\/} property one also has to exclude 
all stationary perturbations other than the Kerr mode. 
While the stationary perturbations do not need to be normalizable, 
they are, however, subject to certain boundary conditions following 
from asymptotic flatness and regularity requirements. 
Both stationary and dynamical perturbations must be
analyzed separately in the sectors $\ell > 1$ and $\ell = 1$.

The vacuum perturbations with odd parity are obtained from
Eq. (\ref{GE-9}), which yields
\be
\frac{1}{R^2} \cod \left[ R^4 \diff
\left( \frac{H}{R^2} \right) \right] + 
\lambda \frac{H}{R^2} = 0 .
\label{US-1}
\ee
This equation holds for all values of $\ell$ and comprises 
the complete information. The usual way to derive the 
RW equation from Eq. (\ref{US-1}) is to decompose 
the one-from $H$ with respect to Schwarzschild coordinates, 
$H = H_{t} \diff t + H_{r} \diff r$, and to use the integrability 
condition to eliminate $H_{t}$. This yields an equation for 
$H_{r}$ alone, which is then cast into a wave equation for the
function $(1-2M/r) H_{r}/r$. 
This can also be achieved in a coordinate-invariant way as follows:
Using the integrability condition $\cod H = 0$ to introduce the 
scalar potential $\Phi$ according to $H = \tilast \diff (R \Phi)$,
one may integrate Eq. (\ref{US-1}). This yields Eq. (\ref{US-3}) 
below for the potential $\Phi$ instead of $\Psi$.  

Here we proceed in a different way, which is also 
coordinate-invariant. The basic observation is that
in two dimensions the field strength two-form assigned to a one-form
is equivalent to a scalar field. We therefore introduce the scalar 
field $\Psi$ according to
\bdm
\Psi \equiv R^3 \tilast \diff \left( \frac{H}{R^2} \right),
\edm
where the factor $R^3$ turns out to be convenient. Applying the 
operator $\tilast \diff$ on Eq. (\ref{US-1}) and using the above 
definition yields the wave equation
\be
\left[ -\laptil + R \laptil \left( \frac{1}{R}\right) +
\frac{\lambda}{R^2} \right] \Psi = 0,
\label{US-3}
\ee
where the two-dimensional Laplacian of a function is
$\laptil \Psi \equiv - \cod \diff \Psi$, and where we have used
$\tilast \diff \cod = \cod \diff \tilast$. Equation (\ref{US-3}) is
the coordinate-invariant version of the RW equation. In fact, it 
generalizes the RW equation, since it is not restricted to 
perturbations of static background configurations.
(The fact that the RW function 
$\Psi \equiv R^3 \tilast \diff (R^{-2} H)$ and the scalar potential
$\Phi$, defined by $H = \tilast \diff (R \Phi)$, satisfy the same 
equation will be explained at the end of Sect. \ref{section-RN-2}.)

The positivity of the RW potential for $\ell \neq 1$ follows from 
the general expression (\ref{GE-1b}) for $G_{ab}$, which yields
the coordinate-independent vacuum background equation 
$R \laptil R + \sprod{\diff R}{\diff R} = 1$. By virtue of this,
Eq. (\ref{US-3}) assumes the form
\bdm
\left[ -N \laptil + V_{RW} \right] \Psi = 0,
\edm
with
\bdm
V_{RW} \equiv \frac{N}{R^2}\left[
3(N-1)+\ell (\ell + 1) \right] 
\edm
and
$N \equiv \sprod{\diff R}{\diff R}$.
Hence, $V_{RW}$ is positive
for finite values of $R$ if $\diff R$ is space-like 
and $\ell \geq 2$.

We may now use standard Schwarzschild coordinates $r$ and $t$, 
defined by
\be
R(r,t) = r, \; \; \; 
\gtiltens = -N S^{2} \diff t^{2} + \frac{1}{N} \diff r^{2} , 
\label{Schwschild}
\ee
to cast the RW equation into its well-known form. For a Schwarzschild 
background with mass $M$ we have $N(r) = 1-2M/r$, $S(r) = 1$, 
$N \laptil = -\p_t^2 + N \p_r N \p_r$, and thus
\be
\left[\frac{\p^2}{\p t^2} - \frac{\p^2}{\p r_{\star}^2} +
\frac{N}{r^{2}}
\left( \ell (\ell + 1) - \frac{6M}{r} \right) \right] \Psi = 0 ,
\label{US-5}
\ee
with $\diff r_{\star} \equiv N^{-1} \diff r$.
For $\ell \geq 2$ the potential is non-negative in the domain
of outer communications, and vanishes only 
asymptotically. Therefore, Eq. (\ref{US-5}) admits no unstable 
dynamical modes. Furthermore, well-behaved stationary modes with 
$\ell \geq 2$ can also be excluded in a rigorous manner by applying 
the argument given in Appendix \ref{App-B}.

It remains to discuss the perturbations with $\ell = 1$, for
which  Eq. (\ref{US-3}) is immediately seen to admit the 
solution $1/R$. Since $\lambda = 0$, we may also directly
integrate Eq. (\ref{US-1}), which yields
\bdm
\diff \left( \frac{H}{R^{2}} \right) = a 
\frac{6M}{R^{4}} \tilast 1 ,
\edm
where $6 a M$ is a constant of integration. At this point it is 
important to recall that for $\ell = 1$ the one-form 
$H \equiv h$ is {\it not\/} coordinate-invariant, but transforms 
according to $H \rightarrow H + R^2 \diff (f/R^2)$. This implies 
that the solution of the homogeneous part of the above equation
is a pure gauge. Hence, with respect to Schwarzschild coordinates, 
the only admissible solution of the perturbation equations
(stationary and non-stationary) is $H = 2 a (M/ r) \diff t$.
Using $S_{\vartheta}^{\ell = 1} = 0$
and $S_{\varphi}^{\ell = 1} = -\sin^{2}\!\vartheta$,
one finds with Eq. (\ref{GE-2})
\bdm
\delta g_{t \varphi} = -a \, \frac{2M}{r} \,
\sin^{2}\!\vartheta \, ,
\edm
which describes the Kerr metric in first order of the rotation
parameter $a$. In conclusion, we have established the well-known 
result that the only physically admissible odd-parity 
{\it vacuum\/} perturbation of the Schwarzschild metric lies in 
the sector $\ell = 1$ and describes the stationary Kerr mode.

\section{Perturbations of Yang-Mills fields}
\label{section-YME}

We are interested in perturbations 
of spherically symmetric EYM 
solitons and black holes which give rise to odd-parity metric 
excitations. Before deriving the gauge- and coordinate-invariant
expressions for the stress-energy tensor and the YM equations, we 
briefly recall some features of the background configurations.

\subsection{Einstein-Yand-Mills Background configurations}
\label{section-YME-BG}

The spherically symmetric EYM background configurations 
are assumed to be purely 
magnetic \cite{electric-abelian}, but not necessarily static. 
The metric is given by Eq. (\ref{GE-1a}), while the gauge 
potential is parametrized in terms of a scalar field 
$w(x^{b})$ on $\tilde{M}$,
\be
A = (1-w) \hatast \diff \taur \, ,
\label{YMBG-1}
\ee
where $\taur \equiv \ul{\tau} \cdot \ul{e}_{r}$. Here the
$\tau_{k} \equiv \sigma_{k}/(2i)$ are the su(2) generators,
$\ul{e}_{r}$ is the radial unit vector in $\RR^{3}$, and the
$\sigma_{k}$ are the constant Cartesian Pauli-matrices.
The total exterior derivative of the vector valued 
function $\ul{e}_{r}$ is $\hat{\theta}^{A} \ul{e}_{A}$
(with $A = \vartheta, \varphi$), implying that
\bdm
\diff \taur =
\tautheta \diff\vartheta + 
\tauphi \sin \! \vartheta \diff\varphi.
\edm
(See Appendix \ref{App-D} for details.) 
Since $\taur$ is an eigenfunction of the spherical Laplacian
$\diff \hatast \diff \taur = -2 \taur \diff \Omega$, the background 
field strength, $F = \diff A + A \wedge A$, becomes
\be
F = - \diff w \we \hatast \diff \taur + 
(w^{2}-1) \taur \diff \Omega \, .
\label{YMBG-2}
\ee
Using this expression, the components of the stress-energy tensor, 
$T_{\mu \nu} = \frac{1}{4\pi} 
\Trace{F_{\mu \alpha} F_{\nu}^{\; \alpha} - \frac{1}{4}
g_{\mu \nu} F_{\alpha \beta} F^{\alpha \beta}}$, with
respect to the background metric (\ref{GE-1a}) become
\bea
T_{ab} & = & \frac{1}{4 \pi R^{2}} \left[
2 w_{a} w_{b} -\frac{1}{2} \gtil_{ab} \left(
2 w_{c} w^{c}  + \frac{(w^{2}-1)^{2}}{R^{2}}
\right) \right] ,
\nonumber\\
T_{AB} & = & \frac{1}{4 \pi R^{2}} \, g_{AB}
\frac{(w^{2}-1)^{2}}{2 R^{2}} \, , \; \; \; 
T_{Ab} \, = \, 0 \, ,
\label{YMBG-3}
\eea
where $w_{a} \equiv \nabtil_{a} w$, and
where $\Trace{\;}$ denotes the normalized trace,
$\Trace{\tau_{i}^{2}} = 1$.

The background YM equation,
$\Diff \ast F  \equiv \diff \ast F + [A, \ast F] = 0$, 
is obtained from the expression
$\ast F = -\tilast \diff w \we \diff \taur + 
R^{-2} (w^{2}-1)\taur \tilast 1$, using the fact
that $\diff \taur$ commutes with $\ast \diff \taur$, 
and $[\diff \taur, \taur] = \ast \diff \taur$.
One finds
\be
\laptil w = w\, \frac{w^{2}-1}{R^{2}} \, , 
\label{YMBG-4}
\ee
where $\cod = \tilast \diff \tilast$, and 
$\laptil w = - \cod \diff w = \nabtil^{a} \nabtil_{a} w$.
The Einstein equations,
$G_{\mu \nu} = 8 \pi G T_{\mu \nu}$,
are obtained
from the formulas (\ref{GE-1b}) and (\ref{YMBG-3}). 
Also using $T^{\mu}_{\; \mu}=0$, one finds
\bea
&\frac{1}{2} \gtil_{ab} \tilde{R} -\frac{2}{R} \nabtil_{a}
\nabtil_{b} R = \nonumber\\
&G \, \frac{2}{R^{2}} \left[
2 w_{a} w_{b} -\frac{1}{2} \gtil_{ab} \left(
2 w_{c} w^{c}  + \frac{(w^{2}-1)^{2}}{R^{2}}
\right) \right] , \label{YMBG-5} \\
&1 - \frac{1}{2} \laptil (R^{2}) = G \, 
\frac{(w^{2}-1)^{2}}{R^{2}} \, .
\label{YMBG-6}
\eea
Equations (\ref{YMBG-4})  - (\ref{YMBG-6}) are the
spherically symmetric EYM equations in coordinate-invariant 
form. In the static case we may evaluate 
these expressions for the metric (\ref{Schwschild}), 
which yields (a prime denoting the derivative with 
respect to $r$)
\be
\frac{1}{S} \left(N S w' \right)' = w \frac{w^{2}-1}{r^{2}}
\label{BK1}
\ee
for the YM equation (\ref{YMBG-4}), and, with 
$N(r) \equiv 1 - 2 m(r)/r$,
\bea
m' & = & \frac{G}{2} \left[
\frac{(w^{2}-1)^{2}}{r^{2}} + 2 N (w')^{2} \right] ,
\label{BK2}\\
\frac{S'}{S} & = & 2 G \, \frac{(w')^{2}}{r}
\label{BK3}
\eea
for Eq. (\ref{YMBG-6}) and for the trace-free part of 
Eq. (\ref{YMBG-5}), respectively. Two special Abelian solutions 
to Eqs. (\ref{BK1}) - (\ref{BK3}) are the Schwarzschild 
metric, $m(r) = M = \mbox{constant}$, $S =1$, $w=1$, and the 
RN metric with mass $M$ and unit magnetic 
charge, $N = 1 -2 M/r + G/r^{2}$, $S =1$, $w=0$.

Asymptotically flat {\it non-Abelian\/} solutions with finite energy 
and nontrivial gauge fields are the solitons found by Bartnik 
and McKinnon \cite{Bartnik88}, and the corresponding black 
holes with hair \cite{VolkGal89}. They are obtained by numerical
methods and by analyzing the local solutions at the 
singular points of Eqs. (\ref{BK1})-(\ref{BK3}), that is, 
at the origin, $r = 0$, the horizon, $N(r_H)=0$, and at infinity,
$r = \infty$. The local background solutions are given in 
Appendix \ref{App-C}, since their behavior will be crucial to the 
existence of regular singular points of the perturbation equations.

\subsection{Gauge- and coordinate-invariant Yang-Mills perturbations}
\label{section-YME-GC}

In Appendix \ref{App-D} we construct a convenient basis of 
su(2)-valued spherical harmonic one-forms. The odd-parity 
perturbations of the YM potential are then given in 
terms of two one-forms, $\alpha$ and $\beta$, and three 
scalar fields, $\mu$, $\nu$ and $\sigma$, over $\tilde{M}$,
\be
\delta A^{(\ell > 1)} = X_{1} \alpha + X_{2} \beta + 
\mu \taur \diff Y + \nu Y \diff \taur + \sigma \nabhat X_{2}, 
\label{YMP-3}
\ee
where $X_1$, $X_2$, and $X_3$ are a scalar basis of su(2)-valued
spherical harmonics,
\bdm
X_{1} = Y \, \taur , \; \; \; 
X_{2} = \ghat^{AB} \tau_{A} \nabhat_{B} Y , \; \; \; 
X_{3} = \hat{\eta}^{AB} \tau_{A} \nabhat_{B} Y,
\edm
while $Y \equiv Y^{\ell m}$ denote the ordinary spherical 
harmonics. (The antisymmetric tensor $\hat{\eta}_{AB}$ is defined 
by $\hatast \hat{\theta}^A = \hat{\eta}^A_{\, B} \hat{\theta}^B$.)
As usual, the cases $\ell = 1$ and $\ell = 0$ must 
be treated separately: For $\ell = 1$, one has 
$\nabhat X_{2}^{(\ell = 1)} = -Y^{(\ell = 1)} \diff \taur$, 
implying that $\nu$ and $\sigma$ combine to a single amplitude. Hence,
\be
\delta A^{(\ell = 1)} = X_{1} \alpha + X_{2} \beta + 
\mu \taur \diff Y + \nu Y \diff \taur .
\label{YMP-4}
\ee
In contrast to the gravitational sector, the odd-parity YM 
sector is not empty for $\ell = 0$. As $Y^{(\ell=0)}$ is constant, 
$\delta A$ is parametrized in terms of the one-form $\alpha$ and
the function $\nu$,
\be
\delta A^{(\ell = 0)} = \taur \alpha + \nu \diff \taur .
\label{YMP-4bis}
\ee

One may now study the behavior of $\delta A$ under gauge 
transformations, $\delta A \rightarrow \delta A + \Diff \chi$, 
and under coordinate transformations,
$\delta A \rightarrow \delta A + {\cal L}_{X} A$.
Here $\Diff$ is the gauge covariant derivative with respect to 
the background connection (\ref{YMBG-1}), $\chi$ is an su(2)-valued
scalar field with odd parity, and ${\cal L}_{X}$ is the Lie 
derivative with respect to the infinitesimal vector field $X$ 
defined in Eq. (\ref{GE-3}). Considering both gauge and coordinate 
transformations, the following results are established in 
Appendix \ref{App-E}:

For $\ell > 1$ the metric perturbations are originally parametrized 
in terms of the function $\kappa$ and the one-form $h$, while the 
YM amplitudes are given by two one-forms, $\alpha$ and 
$\beta$, and three functions, $\mu$, $\nu$ and $\sigma$. Using the
complete gauge and coordinate freedom, the entire set of perturbations
reduces to three one-forms, $H$, $A$ and $B$, and one function, $C$,
all of which are invariant under both coordinate and gauge 
transformations. Adopting the ODG ($\kappa = 0$) and the YM 
gauge $\mu = \sigma = 0$, the quantities $H$, $A$, $B$, and $C$,
coincide with the original amplitudes $h$, $\alpha$, $\beta$, and 
$\nu$. [See Eqs. (\ref{YMP-9}) and (\ref{Lie-5}).] Hence, all 
physically relevant perturbations with $\ell > 1$ are given by
\bea
\delta g^{(\ell>1)}_{AB} &=&  \delta g^{(\ell>1)}_{ab} = 0 , \;\;\; 
\delta g^{(\ell>1)}_{Ab} = H_{b} S_{A} , 
\nonumber\\
\delta A^{(\ell > 1)} &=& X_{1} A + X_{2} B + C \, Y \diff \taur ,
\label{invpertset2}
\eea
with gauge- and coordinate-invariant amplitudes $H$, $A$, $B$, and $C$. 
The ODG for the metric perturbations, together with the YM 
gauge $\mu = \sigma = 0$ will be called the off-diagonal standard 
gauge (ODSG) henceforth. {\it In the ODSG all gravitational and 
YM perturbations coincide with the corresponding coordinate- 
and gauge-invariant quantities\/}.

For $\ell =1$ the metric perturbations are already off-diagonal and 
there exists a gauge for which the YM scalars $\mu$ and $\nu$ 
vanish, and the remaining amplitudes, $\alpha$ and $\beta$, coincide
with the two gauge-invariant one-forms $a$ and $b$, defined in 
Eq. (\ref{YMP-8}). The perturbations are therefore given by
\bea
\delta g^{(\ell=1)}_{AB} &=&  \delta g^{(\ell=1)}_{ab} = 0 , \;\;\; 
\delta g^{(\ell=1)}_{Ab} = h_{b} S_{A} , 
\nonumber\\
\delta A^{(\ell = 1)} &=& X_{1} a + X_{2} b , 
\label{invpertset1}
\eea
where $a$ and $b$ are gauge-invariant, but neither the metric nor 
the YM perturbations are invariant under coordinate 
transformations. The linearized EYM equations involve, however,
only the gauge- {\it and\/} coordinate-invariant combinations
\be 
\bar{a} \equiv a + \frac{h}{R^{2}}, \; \; \; 
\bar{b} \equiv b + w \frac{h}{R^{2}},
\label{l=1invar}
\ee
and $\diff (R^{-2} h)$, as we shall see later.

For $\ell = 0$ there exist no metric perturbations in the 
odd-parity sector, and the YM perturbations are
comprised within a single gauge-invariant one-form $a$,
defined in Eq. (\ref{1-ell-0}),
\be
\delta g^{(\ell=0)}_{\mu \nu} = 0 , \; \; \; 
\delta A^{(\ell = 0)} = \taur \, a .
\label{l=0gauge}
\ee

\section{The Perturbation Equations}
\label{section-PEQ}

In this section we give the equations governing
the odd-parity perturbations of a spherically
symmetric soliton or black hole EYM background configuration. 
The amplitudes
are parametrized in terms of the gravitational one-form
$H$, the YM one-forms $A$, $B$, and the YM scalar $C$.
All amplitudes are gauge- and coordinate-invariant and,
as we are not introducing specific coordinates, the
resulting equations are not limited to static background
configurations. The derivations are considerably simplified
by adopting the ODSG and by taking advantage of the
su(2) harmonics constructed in Appendix \ref{App-D}.
However, as the computations are still lengthy, we discuss
only the basic steps in sections \ref{section-PEQ-g1}, 
\ref{section-PEQ-e1} and \ref{section-PEQ-e0} for $\ell > 1$, 
$\ell = 1$ and $\ell = 0$, respectively, and give a self-contained 
compilation of the results in Sect. \ref{section-PEQ-su}.

\subsection{Equations for $\ell > 1$}
\label{section-PEQ-g1}

For $\ell > 1$ we may proceed in the ODSG for which the metric and
the YM perturbations coincide with the gauge- and coordinate-invariant
amplitudes $H$, $A$, $B$, and $C$:
\bea
\delta g^{(\ell>1)}_{Ab} & = & H_{b} S_{A} ,
\nonumber\\
\delta A^{(\ell > 1)} & = & X_{1} A + X_{2} B + C \, Y \diff \taur .
\label{l>1g-and-A}
\eea

We start by computing the coordinate-invariant
stress-energy tensor.
According to Eqs. (\ref{GE-6a}) and (\ref{GE-6b}) we have
\be
\delta T_{ab}^{inv} = \delta T_{ab}^{ODG} ,
\; \; \; 
\delta T_{AB}^{inv} = \delta T_{AB}^{ODG} ,
\label{deltaT-1}
\ee
and
\be
\delta T_{Ab}^{inv} =
\delta T_{Ab}^{ODG} - H_{b} \, T^{B}_{\; A} S_{B} ,
\label{deltaT-2}
\ee
since $\kappa = 0$ and $H_{a}=h_{a}$ in the ODG.
The $\delta T_{\mu \nu}$ consist of
perturbations arising from variations with respect
to the metric and the YM fields, 
$\delta T_{\mu \nu} = \delta_{g} T_{\mu \nu} +  
\delta_{A} T_{\mu \nu}$, where
\bea
\delta_{g} T_{\mu \nu} & = & - \frac{1}{4\pi} \mbox{Tr} \Big\{
\frac{1}{4} F_{\alpha \beta} F^{\alpha \beta} 
\delta g_{\mu \nu}
\nonumber\\ & + &
\left(F^{\alpha}_{\; \mu} F^{\beta}_{\; \nu} - \frac{1}{2}
g_{\mu \nu} F^{\alpha}_{\; \gamma} F^{\beta \gamma}
\right) \delta g_{\alpha \beta} \Big\} \nonumber
\eea
and
\bdm
\delta_{A} T_{\mu \nu} = \frac{1}{4\pi} \mbox{Tr} \Big\{
F^{\alpha}_{\; \nu} \delta F_{\alpha \mu} +
F^{\alpha}_{\; \mu} \delta F_{\alpha \nu} -
\frac{1}{2} g_{\mu \nu} 
F^{\alpha \beta} \delta F_{\alpha \beta} \Big\}.
\edm

In the ODSG the linearized field strength,
$\delta F = \Diff \delta A$, is obtained from the 
formula (\ref{l>1g-and-A}) for $\delta A^{(\ell > 1)}$.
Recalling that $\Diff$ is the gauge covariant derivative with
respect to the background potential (\ref{YMBG-1}),
one finds, also using the identities (\ref{XID-1}),
\bea
\delta F^{(\ell > 1)}  & = & 
X_{1} \diff A + X_{2} \diff B -
X_{3} \, C \, \diff \Omega -
B \we \nabhat X_{2} 
\nonumber\\
& + &
(wB-A) \we \taur \diff Y + 
(\diff C - wA) \we Y \diff \taur .
\label{YMP-4-F}
\eea
Using this, as well as the expression (\ref{YMBG-2}) 
for the background field strength $F$ and the 
formulas (\ref{l>1g-and-A}) for the
metric perturbations, we end up with
\be
\delta T_{ab}^{inv} = 0 , \; \; \; 
\delta T_{AB}^{inv} = \frac{1}{4\pi} 
\sprod{B}{\diff w} \, 2 \hat{\nabla}_{\{A} S_{B\}} ,
\label{deltaT-4}
\ee
and
\bea
& &   \delta T_{Ab}^{inv} \nonumber\\
& = & \frac{S_A}{4\pi R^{2}} 
\Big[ (w^{2}-1)(A_{b}-wB_{b}) + R^{2} (\diff B)_{ba}w^{a} -
C w_{b} \nonumber\\
& + & \sprod{H}{\diff w} w_{b} - 
\left( \sprod{\diff w}{\diff w} + 
\frac{(w^{2}-1)^{2}}{R^{2}} \right) H_{b} \Big],
\label{deltaT-5}
\eea
where we recall that all amplitudes are gauge- and
coordinate-invariant. Here and in the following we
use the obvious notations 
$w^{a} \equiv \tilde{g}^{ab} \tilde{\nabla}_{b} w$ and
$\sprod{}{}$ for the inner product with respect to $\gtiltens$, e.g., 
$\sprod{H}{\diff w} \equiv \tilde{g}^{ab} H_{a} \tilde{\nabla}_{b} w$.
[There is no factor $1/2$ in front of the last term in 
Eq. (\ref{deltaT-5}), since, according to Eqs. (\ref{YMBG-3}) 
and (\ref{deltaT-2}), $\delta T_{Ab}^{inv}$ and
$\delta T_{Ab}$ differ by the term
$(8\pi)^{-1} R^{-4} (w^{2}-1)^{2} H_{b}S_{A}$ in the ODG.]

The Einstein equations, 
$\delta G_{\mu \nu}^{inv} = 8 \pi G \delta T_{\mu \nu}^{inv}$,
are now obtained from the above expressions and the
formulae (\ref{GE-9b}) and (\ref{GE-9}) for 
$\delta G_{\mu \nu}^{inv}$. We have already argued that
the $(AB)$-equation,
\be
\cod H = -4 G \, \sprod{B}{\diff w},
\label{E1}
\ee 
is a consequence of the $(Ab)$-equations and the linearized
Bianchi identity. While this was obvious for vacuum perturbations,
one now needs the YM equations given below to verify this fact. 
Hence, the only independent Einstein equation is the
one for the coordinate-invariant metric one-form $H$,
\bea
& & \cod \left[ R^{4} \diff \left( \frac{H}{R^{2}} \right) \right] +
\lambda H \nonumber\\
& = &
4G (w^{2}-1) \left[A - w B - \frac{w^{2}-1}{R^{2}} H \right] \nonumber\\ 
& + &
4G \left[\tilde{\ast} ( R^{2} \diff B + \diff w \we H ) 
\tilde{\ast} \diff w - C \diff w \right] ,
\label{E2}
\eea
where we also recall
that $\lambda \equiv (\ell - 1)(\ell + 2)$.
Here we have used the identities
$\sprod{H}{\diff w} \diff w$ $-$ $\sprod{\diff w}{\diff w} H$ $=$
$\tilde{\ast} (\diff w \we H) \tilde{\ast} \diff w$ and 
$(\diff B)_{ab} w^{b} \diff x^{a}$ $=$ 
$(\tilde{\ast} \diff B) \tilde{\ast} \diff w$.

The linearized YM equations also involve perturbations of both
YM and metric fields. The latter arise from the variation of the
Hodge dual in 
$\delta (\Diff \ast F) = 0$, 
and yield the terms on the RHS of the following general expression:
\bdm
\Diff \ast \delta F + [\delta A,\ast F] = 
2 \, \Diff \ast {\cal F} - \diff \left( 
\frac{\delta \sqrt{-\tilde{g}}}{\sqrt{-\tilde{g}}} \right)
\we \ast F ,
\edm
where ${\cal F}_{\mu \nu} \equiv F_{[\mu}^{\; \, \sigma} 
\delta g_{\nu ] \sigma}$. Since the dual of this is an equation 
between one-forms, and since the odd-parity basis of one-forms is 
five-dimensional for $\ell > 1$, we obtain five equations. 
Again, the computation is considerably simplified
in the ODSG for which we may use the gauge-invariant perturbations 
given in Eqs. (\ref{l>1g-and-A}). As expected, it turns out that 
two YM equations can be obtained from the remaining ones.
Using the tools developed in Appendix \ref{App-D}, we  
eventually end up with the following set of equations for 
the one-forms $A$, $B$ and the scalar $C$:
\bea
& & \cod \left( R^{2} \diff A \right) + 
[\lambda + 2 ( w^{2} + 1)] A
-2 [\lambda +2] w B
\nonumber\\
& - & 2 w \diff C + 2 C \diff w = 
(\lambda + 2) \frac{w^{2}-1}{R^{2}} H ,
\label{delYM-1}
\eea
\bea
& & \cod \left( R^{2} \diff B \right) - 
2 w A  + [\lambda + (w^{2} +1)] B + \diff C
\nonumber\\
& = & \cod(H \we \diff w) - w \frac{w^{2}-1}{R^{2}} H ,
\label{delYM-2}
\eea
\be
C = R^{2} \cod B - \sprod{\diff w}{H} .
\label{delYM-3}
\ee
The remaining two YM equations are the integrability conditions for 
Eqs. (\ref{delYM-1}) and (\ref{delYM-2}).
Also using Eq. (\ref{delYM-3}), these become
\be
\cod \left[ A + \frac{1-w^{2}}{R^{2}} H \right] = -
2 \sprod{B}{\diff w} ,
\label{YMint-1}
\ee
and
\bea
& & \laptil C - [\lambda + (w^{2} +1)] \frac{C}{R^{2}} 
\nonumber\\ 
& = &  2 \sprod{A}{\diff w} - w \cod A + [\lambda + 2] 
\frac{\sprod{\diff w}{H}}{R^{2}}.
\label{YMint-2}
\eea

Since Eqs. (\ref{E1}), (\ref{YMint-1}) and (\ref{YMint-2})
are consequences of the remaining equations, the complete system
of perturbation equations consists of the three 
coupled equations (\ref{E2}), (\ref{delYM-1}) and (\ref{delYM-2}) 
for the three gauge- and coordinate-invariant one-forms $A$, $B$ 
and $H$, where $C$ is given by Eq. (\ref{delYM-3}). It will also 
turn out to be convenient to write these equations in terms of 
the one-forms $\bar{A}$ and $\bar{B}$, defined by
\be
\bar{A} \equiv A  + \frac{H}{R^{2}} , \; \; \; 
\bar{B} \equiv B  + w \frac{H}{R^{2}} ,
\label{bars}
\ee
in terms of which Eqs. (\ref{E2}), (\ref{delYM-1}) and 
(\ref{delYM-2}) assume the form 
\bea
& & \cod \left( R^{4} F_{H} \right) +  \lambda H 
\nonumber\\
& = &  4 G \left[ ( R^{2} \tilde{\ast} F_{B} ) \tilde{\ast} \diff w -
C \diff w + (w^{2}-1) (\bar{A}-w \bar{B}) \right], 
\label{EYM-01}
\eea
\bea
& & \cod \left( R^{2} F_{A} \right) +
\lambda \left(\bar{A} - 2w \bar{B} + w^{2} \frac{H}{R^{2}} \right) 
\nonumber\\
& = &  - 2 (w^{2}+1) \bar{A} + 4 w \bar{B} - 2 C \diff w + 2 w \diff C, 
\label{EYM-02}
\eea
\bea
& &\cod \left( R^{2} F_{B} \right) +
\lambda \left( \bar{B} - w \frac{H}{R^{2}} \right) 
\nonumber\\
& = &
2 w \bar{A} - (w^{2}+1) \bar{B} - \diff C ,
\label{EYM-03}
\eea
with
\be
C = R^{2} \left[ \cod \bar{B} - 
w \cod \left( \frac{H}{R^{2}} \right) \right] .
\label{EYM-04}
\ee
Here we have introduced the two-forms 
$F_{A}$, $F_{B}$ and $F_{H}$, which are defined 
in terms of $H$, $\bar{A}$ and $\bar{B}$ as follows:
\bea
&&
F_{A} \equiv \diff \bar{A} - F_{H} , \; \; \; 
F_{B} \equiv \diff \bar{B} - w F_{H} ,
\nonumber\\
&&
F_{H} \equiv \diff \left(\frac{H}{R^{2}}\right) ,
\label{twoforms} 
\eea
i.e., $F_{A} = \diff A$, $F_{B} = \diff B + \diff w \we R^{-2}H$.
The three equations (\ref{EYM-01}) -(\ref{EYM-03}) for the invariant 
one-forms $\bar{A}$, $\bar{B}$ and $H$, with $C$ according to 
Eq. (\ref{EYM-04}), govern all physical odd-parity perturbations 
with $\ell > 1$. We shall now argue that these equations 
hold for $\ell =1$ as well, provided that one sets $C=0$.

\subsection{Equations for $\ell = 1$}
\label{section-PEQ-e1}

For $\ell = 1$ the metric perturbations are off-diagonal and
described by the one-form $h$, while the YM potential 
is parametrized in terms of two one-forms $a$ and $b$,
\be
\delta g^{(\ell=1)}_{Ab} = h_{b} S_{A} , \; \; \; 
\delta A^{(\ell = 1)} = X_{1} a + X_{2} b .
\label{l=1g-and-A}
\ee
Although $a$ and $b$ are gauge-invariant, they are not invariant 
under coordinate transformations, and neither is $h$. As the 
linearized YM and Einstein equations are coordinate-invariant, 
these will only involve the gauge- and coordinate-invariant one 
forms $\bar{a}$ and $\bar{b}$ defined in Eq. (\ref{l=1invar}). 

The perturbation equations for $\ell =1$
are obtained from the equations for $\ell > 1$ as follows:
The linearized field strength two-form,
$\delta F = \Diff \delta A$, for 
the background potential (\ref{YMBG-1}) and the perturbation
(\ref{l=1g-and-A}) becomes
\bea
\delta F^{(\ell = 1)} & = & X_{1} \diff a + X_{2} \diff b 
\nonumber\\
& + & (wb-a) \we \taur \diff Y + (b-wa) \we Y \diff \taur \, .
\label{YMP-4-F1}
\eea
Formally, this is also obtained from the expression
(\ref{YMP-4-F}) for $\delta F^{(\ell > 1)}$ by substituting
$a$ for $A$, $b$ for $B$ and by setting $C=0$, where one also 
has to use $\nabhat X_{2}^{(\ell = 1)} = - Y^{(\ell = 1)} \diff \taur$; 
see Appendix \ref{App-D} for details. Hence, the invariant stress 
energy tensor for $\ell = 1$ is obtained from the expressions 
(\ref{deltaT-4}) and (\ref{deltaT-5}) for $\ell > 1$ by applying 
these substitutions and by replacing $h$ for $H$. This yields
\bdm
\delta T_{ab}^{inv} = 0 , \;\;\;  \delta T_{AB}^{inv} = 0
\edm
and
\bea
\delta T_{Ab}^{inv} \diff x^{b} & = & \frac{S_A}{4\pi R^{2}} 
\left[
(w^{2}-1)(a-wb) + R^{2} (\tilde{\ast} \diff b) 
\tilde{\ast} \diff w \right] 
\nonumber\\
& + & \frac{S_A}{4\pi R^{2}} 
\left[ \tilde{\ast} (\diff w \we h) \tilde{\ast} \diff w -
\frac{(w^{2}-1)^{2}}{R^{2}} h \right] S_{A} . \nonumber
\eea
The coordinate-invariance of the last expression becomes
manifest by writing it in terms of the one forms 
$\bar{a}$ and $\bar{b}$ given in Eq. (\ref{l=1invar}).
One finds 
\bea
\delta T_{Ab}^{inv} \diff x^{b} & = & \frac{S_A}{4\pi R^{2}} 
(w^{2}-1)(\bar{a}-w \bar{b})
\nonumber\\ & + &
\frac{S_A}{4\pi} 
\tilde{\ast} \left[
\diff \bar{b} - w \diff \left(\frac{h}{R^{2}} \right) \right]
\tilde{\ast} \diff w,
\label{deltaT-5bisGI}
\eea
where the metric perturbation enters only via
the invariant two-form
$\diff (R^{-2}h)$.

It is now obvious that the complete set of linearized EYM equations
in terms of the gauge {\it and} coordinate-invariant amplitudes 
$\bar{a}$ and $\bar{b}$ is obtained from 
Eqs. (\ref{EYM-01})-(\ref{EYM-03}), by substituting
$\bar{A} \rightarrow \bar{a}$, $\bar{B} \rightarrow \bar{b}$ and 
$C \rightarrow 0$. As we also have to substitute $H \rightarrow h$,
the LHS of Eqs. (\ref{EYM-01})-(\ref{EYM-03}) would, at a first glance, 
involve the non-coordinate-invariant amplitude $h$. However, since
$\lambda = 0$ for $\ell = 1$, the terms involving $h$ itself vanish 
identically. We also 
point out that the algebraic equation (\ref{EYM-04}) for $C$ is not 
present for $\ell = 1$, because the basis of one-forms is reduced by
one dimension. The complete set of perturbation equations in the 
sector $\ell = 1$ thus assumes the surprisingly simple form
\bea
& \cod \left( R^{4} F_{h} \right) & - 
4 G \left[ ( R^{2} \tilde{\ast} F_{b} ) \tilde{\ast} \diff w +
(w^{2}-1) (\bar{a}-w \bar{b}) \right] = 0 ,
\nonumber\\
&\cod \left( R^{2} F_{a} \right) & + 
2 (w^{2}+1) \bar{a} - 4 w \bar{b} = 0,
\nonumber\\
& \cod \left( R^{2} F_{b} \right) & -
2 w \bar{a} + (w^{2}+1) \bar{b} = 0 ,
\label{EYM-03bis}
\eea
where $\bar{a}$ and $\bar{b}$ are the gauge- and 
coordinate-invariant one-forms given in Eq. (\ref{l=1invar}), in terms
of which the two-forms $F_{h}$, $F_{a}$ and $F_{b}$
are defined by
\bdm
F_{h} \equiv \diff \left(\frac{h}{R^{2}}\right) , \; \; \;
F_{a} \equiv \diff \bar{a} - F_{h} , \; \; \; 
F_{b} \equiv \diff \bar{b} - w F_{h} . 
\edm

\subsection{Equations for $\ell = 0$}
\label{section-PEQ-e0}

For $\ell = 0$ there exist no metric perturbations
with odd parity, and $\delta A$ can be expressed in
terms of a single gauge-invariant one-form
\bdm
\delta g^{(\ell=0)}_{\mu \nu} = 0 , \;\;\; 
\delta A^{(\ell = 0)} = \taur \, a .
\edm
The field strength is obtained by setting $Y=1$,
$B=0$, $C=0$ in the expression (\ref{YMP-4-F}) for 
$\delta F^{(\ell > 1)}$, and by substituting
$a$ for $A$, 
\bdm
\delta F^{(\ell = 0)} = \taur \diff a - w a \we \diff \taur .
\edm
The correct perturbation equation is now obtained from 
Eq. (\ref{EYM-02}) by setting $\bar{B}=C=H=0$, where
$\lambda = -2$ for $\ell = 0$. Also substituting $a$ for 
$\bar{A}$, Eqs. (\ref{EYM-02}) and (\ref{twoforms}) yield
\be
\cod \left(R^{2} F_{a} \right) + 2 w^{2} a = 0 , \;\;\; 
\mbox{with $F_{a} \equiv \diff a$.}
\label{EYM-02bisbis}
\ee

\subsection{Summary}
\label{section-PEQ-su}

All odd-parity perturbations of spherically
symmetric, not necessarily static EYM configurations
are governed by the three equations 
\bea
& & \cod \left( R^{4} F_{H} \right) +  \lambda H 
\nonumber\\
& = & 4 G \left[ ( R^{2} \tilde{\ast} F_{B} ) \tilde{\ast} \diff w -
C \diff w + (w^{2}-1) (\bar{A}-w \bar{B}) \right], 
\label{S-1}
\eea
\bea
& & \cod \left( R^{2} F_{A} \right) +
\lambda \left(\bar{A} - 2w \bar{B} + w^{2} \frac{H}{R^{2}} \right) 
\nonumber\\
& = & - 2 (w^{2}+1) \bar{A} + 4 w \bar{B} - 2 C \diff w + 2 w \diff C, 
\label{S-2}
\eea
\bea
& &\cod \left( R^{2} F_{B} \right) +
\lambda \left( \bar{B} - w \frac{H}{R^{2}} \right) 
\nonumber\\
& = &
2 w \bar{A} - (w^{2}+1) \bar{B} - \diff C ,
\label{S-3}
\eea
for the three gauge- and coordinate-invariant
one-forms $\bar{A}$, $\bar{B}$, and $H$, where
$\lambda \equiv (\ell-1)(\ell+2)$,
\be
C \equiv (1-\delta^{\ell}_{0}-\delta^{\ell}_{1} ) R^{2}
\left[ \cod \bar{B} - 
w \cod \left( \frac{H}{R^{2}} \right) \right] ,
\label{S-4a}
\ee
and
\bdm
F_{H} \equiv \diff \left(\frac{H}{R^{2}}\right), \;\;\;
F_{A} \equiv \diff \bar{A} - F_{H}, \;\;\; 
F_{B} \equiv \diff \bar{B} - w F_{H}. 
\edm
The above equations are valid for all values of $\ell$, where
only Eq. (\ref{S-2}) with $H=\bar{B} = 0$ is present
for $\ell = 0$. However, the expressions for the 
gauge- and coordinate-invariant amplitudes 
in terms of the original metric and YM perturbations are different 
for $\ell > 1$, $\ell = 1$ and $\ell = 0$, respectively; see
Appendix \ref{App-E}.

$\ell >1$: The original metric perturbations 
are described by a one-form
$h$ and a function $\kappa$, while the YM perturbations
are given in terms of two one-forms, $\alpha$ and $\beta$, and
three functions, $\mu$, $\nu$ and $\sigma$:
\bea
&&\delta g_{ab} = 0 , \;\;\; 
\delta g_{Ab} \diff x^{b} = h \, S_{A} , \;\;\; 
\delta g_{AB} = 2 \kappa \nabhat_{\{A} S_{B\}} ,
\nonumber\\
&&\delta A = X_{1} \alpha + X_{2} \beta + 
\mu \taur \diff Y + \nu Y \diff \taur +
\sigma \nabhat X_{2}. 
\nonumber
\eea
In terms of these amplitudes the invariant quantities appearing in
Eqs. (\ref{S-1})-(\ref{S-3}) are, according to Appendix \ref{App-E},
\bea
\bar{A} & \equiv & \alpha + \frac{h}{R^{2}} -\diff
\left(\mu + w \sigma + w^{2} \frac{\kappa}{R^{2}} \right) ,
\nonumber\\
\bar{B} & \equiv & \beta + w \frac{h}{R^{2}} -\diff
\left(\sigma + w \frac{\kappa}{R^{2}} \right) ,
\nonumber\\
H & \equiv & h - R^{2} \diff \left( \frac{\kappa}{R^{2}} \right).
\label{S-7}
\eea

$\ell =1$:
The original metric perturbations are described by the one-form
$h$, while the YM perturbations
are given in terms of two one-forms, $\alpha$ and $\beta$, and
two functions, $\mu$ and $\nu$:
\bea
&&\delta g_{ab} = 0, \;\;\;
\delta g_{Ab} \diff x^{b} = h \, S_{A}, \;\;\; 
\delta g_{AB} = 0,
\nonumber\\
&&\delta A = X_{1} \alpha + X_{2} \beta + 
\mu \taur \diff Y + \nu Y \diff \taur.
\eea
The invariant quantities now are
\bea
\bar{A} & \equiv & \alpha + \frac{h}{R^{2}} -\diff
\left(\frac{\mu - w \nu}{1-w^{2}}\right) ,
\nonumber\\
\bar{B} & \equiv & \beta + w \frac{h}{R^{2}} -\diff
\left(\frac{w\mu -  \nu}{1-w^{2}}\right) ,
\nonumber\\
H & \equiv & h .
\label{S-10}
\eea

$\ell =0$:
There exist no metric perturbations, and the YM perturbations 
are given in terms of a one-form, $\alpha$, and a function, $\nu$: 
\bdm
\delta A = \taur \alpha + \nu \, \diff \taur .
\edm
In terms of $\alpha$ and $\nu$ the invariant quantity 
$\bar{A}$ is given by
\be
\bar{A}  \equiv  \alpha - \diff \left( \frac{\nu}{w} \right) , 
\label{S-12}
\ee
and, as mentioned above, 
the correct perturbation equation is Eq. (\ref{S-2}) with
$\lambda = -2$, $C=0$, $\bar{B} = 0$ and $H = 0$.

\section{Non-Abelian stability and local uniqueness
of the Reissner-Nordstr\"om solution}
\label{section-RN}

For $w \equiv 0$ the static, spherically symmetric EYM equations
(\ref{BK1})-(\ref{BK3}) admit the RN solution with unit magnetic 
charge. The stability and local uniqueness properties of the RN 
metric with respect to non-Abelian perturbations are, therefore, 
obtained from Eqs. (\ref{S-1})-(\ref{S-3}), which decouple into 
two sets for $w \equiv 0$: The first set, involving the one-forms 
$H$ and $\bar{A}$ only, is obtained from Eqs. (\ref{S-1}) and (\ref{S-2}),
\bea
& & \cod \left[ R^{4} \diff \left(\frac{H}{R^{2}} \right) \right] + 
    \lambda H = -4G \bar{A} , 
\nonumber\\
& & \cod \left[ R^{2} \diff \left( \bar{A} - \frac{H}{R^{2}} 
    \right) \right] + (\lambda + 2) \bar{A} = 0 
\; \; \mbox{for $\ell \geq 1$.}
\label{AH-eqs-2}
\eea
Since $w \equiv 0$, the remaining equation for $\bar{B}$ does
not contain the amplitudes $H$ and $\bar{A}$. Using 
$C = R^{2} \cod \bar{B}$ for $\ell >1$ and $C=0$ for
$\ell = 1$, we have
\bea
&&\cod \left( R^{2} \diff \bar{B} \right) +
\diff \left( R^{2} \cod \bar{B} \right) + (\lambda \!+\! 1) \bar{B} = 0
\; \mbox{for $\ell > 1$} ,
\label{B-eq-1}\\
&&\cod \left( R^{2} \diff \bar{B} \right) + \bar{B} = 0
\; \mbox{for $\ell = 1$} .
\label{B-eq-2}
\eea

Since $\bar{A}$ is the gauge- and coordinate-invariant version of
the amplitude in front of the isospin harmonics $\taur Y^{\ell}$,
Eqs. (\ref{AH-eqs-2}) govern the Abelian part of the perturbations,
that is, Einstein-Maxwell perturbations of the RN metric. In
contrast to this, Eqs. (\ref{B-eq-1}) and (\ref{B-eq-2}) for $\bar{B}$ 
are not present in the Abelian case, and describe non-Abelian 
perturbations of the RN metric with $\ell > 1$ and $\ell = 1$, respectively.

\subsection{Perturbations with $\ell > 1$}
\label{section-RN-1}

We start with Eqs. (\ref{AH-eqs-2}) describing the Abelian part 
of the perturbations. For $\ell > 1$ the integrability 
conditions for these equations are $\cod \bar{A} = 0$ and 
$\cod H = 0$, implying the existence of two scalar fields, 
$\Psi_H$ and $\Psi_A$, defined by
\bdm
\tilast \diff \left( R \Psi_H \right) \equiv \sqrt{\lambda} H, 
\;\;\; 
\tilast \diff \Psi_A \equiv \sqrt{4 G} \bar{A} .
\edm
Substituting $\Psi_H$ and $\Psi_A$ for $H$ and $\bar{A}$ in 
Eqs. (\ref{AH-eqs-2}), and integrating both equations yields 
the following coupled wave equations for the scalar fields
$\Psi_H$ and $\Psi_A$:
\bea
\laptil \Psi_H & = & \left[ R \cod \left( \frac{\diff R}{R^2}
\right) + \frac{\lambda}{R^2} \right] \Psi_H + 
\frac{\sqrt{4 G \lambda}}{R^3} \Psi_A ,
\nonumber\\
\laptil \Psi_A & = & 
\frac{\sqrt{4 G \lambda}}{R^3} \Psi_H  + 
\left[ \frac{\lambda + 2}{R^2} + 
\frac{4G}{R^4} \right] \Psi_A  . \nonumber
\eea
For $w=0$ the background equation (\ref{YMBG-6}) becomes
$R^3 \cod (\diff R / R^2) = 3 \sprod{\diff R}{\diff R} -1 + G / R^{2}$.
Using this and introducing standard Schwarzschild coordinates,
$R=r$, $\sprod{\diff R}{\diff R} = N = 1 - 2M / r + G / r^{2}$,
yields 
\bea
\left[ - \laptil + \frac{1}{r^2} \left(
\lambda + 2 - \frac{3M}{r} + \frac{4 G}{r^2} \right) 
\right] & &
\left( \begin{array}{c}
\Psi_H \\ \Psi_A
\end{array} \right)
\nonumber\\ 
+ \frac{1}{r^3}
\left( \begin{array}{cc}
-3M & \sqrt{4 G \lambda} \\
\sqrt{4 G \lambda} & 3M
\end{array} \right) & &
\left( \begin{array}{c}
\Psi_H \\ \Psi_A
\end{array} \right)  = 0 . 
\label{Psi12-eq2}
\eea

The above equation was first obtained by Moncrief by different means
\cite{VM}. Since the off-diagonal part of the potential is symmetric 
and constant, Eq. (\ref{Psi12-eq2}) can be decoupled. Using the 
non-negativity of $N(r)$, as well as the regularity condition
$M \geq G$, the eigenvalues of the potential are found to be
positive, implying the absence of unstable modes. Taking advantage 
of the argument presented in  Appendix \ref{App-B}, {\it stationary\/} 
modes are excluded as well. (The eigenvalues of the potential are 
positive for finite $r$ and behave like 
$\ell (\ell -1) r^{-2} + {\cal O}(r^{-3})$ for $r \rightarrow \infty$, 
implying that the asymptotically finite solutions behave like 
$r^{-\ell}$.) Hence, there exist neither unstable modes nor admissible 
stationary solutions to Eqs. (\ref{AH-eqs-2}) for $\ell > 1$.

In order to discuss the non-Abelian part of the perturbations
we introduce the scalar fields 
$\Pi_{1} \equiv R^{2} \tilast \diff \bar{B}$ and 
$\Pi_{2} \equiv R^{2} \cod \bar{B}$. In terms of these, 
Eq. (\ref{B-eq-1}) assumes the form
\be
\tilast \diff \Pi_{1} + \diff \Pi_{2} + 
(\lambda + 1) \bar{B} = 0,
\label{Hodge-barB}
\ee
which can also be viewed as the Hodge decomposition of the
one-form $\bar{B}$ (see the comments below). Applying the 
operators $\tilast \diff$ and $\cod \equiv \tilast \diff \tilast$ to
this, it is immediately seen that $\Pi_{1}$ and $\Pi_{2}$ are subject to
the same equation, namely
\be
-\laptil \Pi_{i}  + \frac{\lambda + 1}{R^{2}} \Pi_{i} ,
\; \; \; \mbox{$i=1,2$},
\label{Pi-i}
\ee
where we recall that $\lambda = (\ell-1)(\ell+2)$.
With respect to the static RN background, $R(r,t)=r$,
$\gtiltens = N(-\diff t^{2} + \diff r_{\star}^{2})$, with
$N(r) = 1 -2M / r + G / r^{2}$ and 
$\diff r_{\star} = \diff r / N$, one has
\be
\left[ \frac{\p^{2}}{\p t^{2}} - 
\frac{\p^{2}}{\p r_{\star}^{2}} + N(r)
\frac{\ell(\ell + 1) -1}{r^{2}} \right] \Pi_{i}= 0 .
\label{Pi-i2}
\ee
Since the operator is positive, unstable modes are absent. Furthermore, 
well-behaved stationary solutions are excluded as well, since the potential 
is of the type required to apply the argument given in Appendix \ref{App-B}.
[Also note that the one-form $\bar{B}$ is obtained directly from 
$\Pi_{1}$ and $\Pi_{2}$ by Eq. (\ref{Hodge-barB}).]

As we shall continue to use the above method, it is worthwhile 
noticing the following: In two dimensions an arbitrary one-form $\theta$, 
say, gives rise to two scalar fields, $g_{1} \equiv \cod \theta$ and 
$g_{2} \equiv \tilast \diff \theta$. On the other hand, the Hodge 
decomposition of a one-form in two dimensions involves two different 
scalar fields, $\theta \equiv \diff f_{1} + \tilast \diff f_{2}$
(provided that the harmonic part vanishes). If 
$\theta$ is subject to a linear wave equation, then the latter gives 
rise to an {\it algebraic\/} relation between the two 
different parameterizations, although the two scalar pairs 
are defined on different differential levels. (This is also the 
reason why, in Sect. \ref{subsection-GE-LU}, we have 
obtained the same RW equation (\ref{US-3}) for $\Psi$ and $\Phi$, defined 
by  $\Psi = R^3 \tilast \diff (H/R^2)$ and $H = \tilast \diff (R \Phi)$, 
respectively.)

\subsection{Perturbations with $\ell = 1$}
\label{section-RN-2}

Defining $\Pi_{1} \equiv \tilast R^{2} \diff \bar{B}$ as for
$\ell > 1$, Eq. (\ref{B-eq-2}) for $\bar{B}$ reduces to 
$\tilast \diff \Pi_{1} + \bar{B} = 0$. Applying the operator 
$\tilast \diff$ yields the same equation for $\Pi_{1}$ as before, 
that is, Eq. (\ref{Pi-i}), where now $\ell = 1$. As the potential 
remains positive for $\ell = 1$, we conclude that Eq. (\ref{B-eq-2}) 
admits neither unstable modes nor admissible stationary perturbations, 
which establishes the stability and the local uniqueness of the RN 
metric with respect to non-Abelian odd-parity perturbations.

It remains to consider Eqs. (\ref{AH-eqs-2}) for $\ell = 1$, i.e.,
for $\lambda = 0$. As these equations are also present in the
Abelian case, we will recover the absence of unstable modes, while 
the only stationary perturbations are those describing the 
Kerr-Newman excitations of the RN solution. This is seen as follows:
For $\lambda = 0$ the only integrability condition for 
Eqs. (\ref{AH-eqs-2}) is $\cod \bar{A} = 0$. Using this to define
the scalar field $\Psi$ according to
\be
\tilast \diff \Psi \equiv \bar{A},
\label{Psi}
\ee
Eqs. (\ref{AH-eqs-2}) can be integrated, which yields
\bea
& & R^{4} \tilast \diff \left(\frac{H}{R^{2}}\right) + 
4 G  \Psi =  6 M a ,
\label{au1}\\
& & R^{2} \left[ \laptil \Psi +
\tilast \diff \left(\frac{H}{R^{2}}\right) \right]
- 2 \Psi = 0 ,
\label{au2}
\eea
where $6M a$ is a constant of integration, and where we have used 
the fact that $\Psi$ is defined up to a constant in order to neglect 
the second constant of integration. Eliminating the gravitational 
perturbation $H$ from the above equations yields the following 
inhomogeneous wave equation for $\Psi$:
\be
\left[-\laptil + \frac{2}{R^{2}} + \frac{4G}{R^{4}}
\right] \Psi = \frac{6 M a}{R^{4}} .
\label{Psi--eq}
\ee
As the operator on the LHS is positive, we conclude again that there 
are no unstable modes. Using standard Schwarzschild coordinates, 
$R=r$, $N = 1 -2M/r + G/r^{2}$, we have 
$\laptil = -N^{-1} \p_{t}^{2} + \p_{r} N \p_{r}$, and the inhomogeneous 
problem admits the particular solution $\Psi = a/r$. By virtue of 
Eq. (\ref{au1}) and definition (\ref{Psi}) this yields, up to a gauge,
\be
H = a(N-1) \diff t, \;\;\; 
\bar{A} = a \frac{N}{r^{2}} \diff t .
\label{res-H-barA}
\ee
Recalling that for $\ell =1$ one has
$\delta g_{a \vartheta = 0}$,
$\delta g_{a \varphi} = - H_{a} \sin^{2} \! \vartheta$ and 
$\delta A = (\bar{A}-H/r^{2}) X_{1} + (\bar{B}-wH/r^{2}) X_{2}$, 
we find with $w=0$ and $\bar{B}=0$
\bea
\delta g_{t \varphi} & = & a \left( \frac{2M}{r} - \frac{G}{r^{2}}
\right) \sin^{2} \! \vartheta  ,
\nonumber\\
\delta A & = & \frac{a}{r^{2}} \taur \cos \! \vartheta \diff t , \nonumber
\eea
which is the Kerr-Newman excitation of the magnetically charged RN 
metric. In order to see this, we compute the {\it electric\/} field,
$\delta E = -\delta F (\p_{t}, \cdot) =
- \Diff \delta A(\p_{t}, \cdot) = 
a \taur  \diff(\cos \! \vartheta / r^{2})$, where we have used 
$D \taur = 0$ and $w = 0$. Hence
\bdm
\delta E = -\taur a  \left( \frac{\sin \! \vartheta}{r^{2}}
\diff \vartheta  + 
\frac{2 \cos \! \vartheta}{r^{3}} \diff r \right).
\edm
Since the magnetic field of the background solution is 
$B = - \taur (\ast \diff \Omega)(\p_{t}, \cdot) =
- \taur (1/r^{2}) \diff r$ 
[see Eq. (\ref{YMBG-2}) for $w=0$], we obtain indeed the 
magnetically charged Kerr-Newman solution in first order of 
the rotation parameter $a$.

\subsection{Perturbations with $\ell = 0$}
\label{section-RN-3}

Since the odd-parity gravitational sector is empty for $\ell = 0$, 
the perturbations of the RN solution are governed by 
Eq. (\ref{S-2}) with $H = \bar{B} = 0$,
$w=0$ and $\lambda = -2$,
\bdm
d \tilast \left( R^{2} \diff \bar{A} \right) = 0.
\edm
With respect to Schwarzschild coordinates
the solution is $\bar{A}= (q / r) \diff t$, where $q$ is
a constant of integration. The perturbation
of the gauge potential now becomes 
$\delta A = \taur (q / r) \diff t$, which
gives rise to a radial electric field,
\bdm
\delta E = -\taur \frac{q}{r^{2}} \diff r .
\edm
Hence, we obtain the embedded magnetic RN solution with
infinitesimal electric charge $q$. (Note that
the metric remains unchanged in first order of $q$.)

In conclusion, we have shown that the RN solution is stable 
with respect to both Abelian {\it and\/} non-Abelian 
odd-parity perturbations for all values of $\ell$. 
Also, the only physically admissible stationary modes are 
the Abelian ones, describing electric Kerr-Newman
($\ell = 1$) and RN ($\ell = 0$) excitations of the
magnetic RN metric.

\section{Non-Abelian stability and local uniqueness
of the Schwarzschild solution}
\label{section-Schw}

The Schwarzschild metric solves the spherically symmetric EYM 
background equations with $w=1$. As the stress-energy tensor 
is quadratic in the field strength, the gravitational perturbations
decouple in first order for all values of $\ell$, and are governed by
the RW equation for vacuum perturbations. The remaining equations, 
describing Abelian and non-Abelian perturbations of the Schwarzschild 
metric, admit no unstable modes, and, for $\ell > 1$, no acceptable 
stationary excitations either. For $\ell = 1$ the only stationary YM
perturbation is the RN mode in the Abelian sector.

\subsection{Perturbations with $\ell > 1$}
\label{section-Schw-A}

The gauge- and coordinate-invariant one-forms $\bar{A}$, $\bar{B}$
and $H$ given in Eqs. (\ref{S-7}) for $\ell > 1$ are well-defined
for $w=1$. The perturbations are, therefore, governed by 
Eqs. (\ref{S-1})-(\ref{S-3}), where Eq. (\ref{S-1}) decouples
for $w=1$ and reduces to the usual equation describing the vacuum 
perturbations of the Schwarzschild metric, 
\be
\cod \left[ R^{4}  \diff \left( \frac{H}{R^{2}} \right)
\right] + \lambda H = 0.
\label{RW-bis}
\ee
In Sect. \ref{subsection-GE-LU} we have already recalled that this 
equation admits neither unstable nor well-behaved stationary solutions 
for $\ell > 1$. 

In order to discuss the non-vacuum perturbations of the Schwarzschild 
metric, it is more convenient to resort to the original one-forms 
$A = \bar{A} - H/R^{2}$ and $B = \bar{B}-w H/R^{2}$, used in 
Sect. \ref{section-PEQ} to derive the perturbation equations. In terms 
of $A$ and $B$, Eqs. (\ref{S-2}) and (\ref{S-3}) become for $w=1$ 
\bea
& & \cod \left(R^{2} \diff A \right) -2 \diff \left(R^{2} \cod B 
\right) + (\lambda + 4) A -2 (\lambda + 2) B = 0 ,
\nonumber\\
& & \cod \left(R^{2} \diff B \right) + \diff \left(R^{2} \cod B 
\right) - 2 A + (\lambda + 2) B = 0 .
\label{AB-Schw}
\eea
The above system is equivalent to four coupled equations for 
four scalar fields. In order to decouple these equations 
completely, we note the following: The terms with $B$ and 
$\cod B$ can be eliminated, which shows that 
the integrability condition is $\cod A = 0$. Using this, and 
applying the co-differential operator on either of the 
above equations, yields a wave equation for the scalar 
field $\cod B$ alone,
\be
\left( -\laptil + \frac{\lambda + 2}{R^{2}} \right)\Pi_{B} = 0, \; \; \; 
\mbox{$\Pi_{B} \equiv R^2 \cod B$}.
\label{Schw1}
\ee
Since the integrability condition implies that the scalar 
$\Pi_{A} \equiv R^2 \cod A$ vanishes, it remains to find the
equations for the field strengths $\diff A$ and $\diff B$, or, 
equivalently, for the scalar fields $\Psi_{A}$ and $\Psi_{B}$, 
defined by
\bdm
\Psi_{A} \equiv \tilast R^{2} \diff A, \; \; \; 
\Psi_{B} \equiv \sqrt{\lambda + 2} \, \tilast R^{2} \diff B .
\edm
Applying the operator $\tilast \diff$ on Eqs. (\ref{AB-Schw}) then
yields the system
\be
\left[ - \laptil + \frac{1}{R^2}
\left( \begin{array}{cc}
\lambda + 4 & -2 \sqrt{\lambda + 2} \\
-2 \sqrt{\lambda + 2} & \lambda + 2
\end{array} \right) \right]
\left( \begin{array}{c}
\Psi_A \\ \Psi_B
\end{array} \right)  = 0 , 
\label{Schw3}
\ee
which can be diagonalized, since the potential is symmetric and constant.
The eigenvalues are
\bdm
\lambda + 3 \pm \sqrt{4 \lambda + 9} = 
\left\{ \begin{array}{l}
(\ell + 1)(\ell + 2)  \\
\ell (\ell-1) 
\end{array} \right. .
\edm
Having solved Eqs. (\ref{Schw1}) and (\ref{Schw3}), the expressions 
for the one-forms $A$ and $B$ in terms of the scalar fields are
obtained from the original equations (\ref{AB-Schw}):
\bea
A & = & -\frac{1}{\lambda} \tilast \diff \left(
\Psi_{A} + \frac{2}{\sqrt{\lambda + 2}} \Psi_{B}
\right) ,
\nonumber\\
B & = & -\frac{1}{\lambda (\lambda + 2)}  \left[ 
\tilast \diff \left(
2 \Psi_{A} + \frac{\lambda + 4}{\sqrt{\lambda + 2}} \Psi_{B}
\right) + \lambda \, \diff \Pi_B \right] .
\nonumber
\eea

Since the operators in Eqs. (\ref{Schw1}) and (\ref{Schw3}) are 
positive, we conclude, using the argument given in 
Appendix \ref{App-B}, that the Schwarzschild solution admits 
neither unstable nor stationary non-Abelian odd-parity modes
with $\ell > 1$.

\subsection{Perturbations with $\ell = 1$}
\label{section-Schw-B}

For $w=1$ Eq. (\ref{S-1}) decouples for all values of $\ell$.
The vacuum perturbations of the Schwarzschild metric with $\ell = 1$ 
are, therefore, governed by Eq. (\ref{RW-bis}) with $\lambda = 0$.
We have already recalled in Sect. \ref{subsection-GE-LU} that this
equation cannot give rise to unstable modes, while it admits
the well-behaved stationary solution $H = (2aM/r)\diff t$, giving
rise to the Kerr excitation of the Schwarzschild metric,
\be
\delta g_{t \varphi} = -a \frac{2M}{r} \sin^{2} \! \vartheta .
\label{Sch1-1}
\ee

In order to analyze the YM sector, we first note that the gauge 
invariant quantities introduced in Eqs. (\ref{YMP-8}) for $\ell = 1$ 
are {\it not\/} well-defined if $w=1$. Hence, the $\ell = 1$
perturbations of the Schwarzschild background require a special
treatment: For $w=1$ and $\ell = 1$ we define $a$, $b$ and $c$ 
in the same way as for $\ell > 1$, that is, by Eqs. (\ref{YMP-9}). 
Hence, $a= \alpha - \diff \mu$, $b= \beta$ and
$c = \nu - \mu$, where the one-forms $\alpha$, $\beta$ and the
scalars $\mu$, $\nu$ parametrize $\delta A^{(\ell = 1)}$
according to Eq. (\ref{YMP-4}). Since the gauge fields vanish
on the background, all YM amplitudes are coordinate-invariant, 
and it remains to consider their behavior under gauge transformations. 
By virtue of Eqs. (\ref{YMP-7a}) and (\ref{YMP-7b}) $c$ remains
invariant, whereas $a$ and $b$ transform according to 
$ \rightarrow a + \diff f_{2}$, $b \rightarrow b + \diff f_{2}$. 
Repeating the arguments given in Sect. \ref{section-PEQ-e1}, the 
perturbation equations for $w=1$ and $\ell = 1$ eventually become
\bea
&& \cod \left( R^{2} \diff a \right) - 2 \diff c + 4 (a-b) = 0,
\nonumber\\
&& \cod \left( R^{2} \diff b \right) +  \diff c - 2 (a-b) = 0,
\nonumber\\
&& R^{2} \cod \left( a-b \right) + c = 0 ,
\nonumber
\eea
where $c$, $\diff a$, $\diff b$ and $(a-b)$ are gauge-invariant.
Subtracting the first from the second equation, and using the 
third one to eliminate $c$, we obtain an equation for the one-form 
$(b-a)$. This is decoupled in the usual way, that is, by introducing 
two scalar fields according to
\bdm
c_{1} \equiv R^{2} \cod (b-a) , \; \; \; 
c_{2} \equiv R^{2}\tilast \diff (b-a) .
\edm
Applying the operators $\cod$ and $\tilast \diff$ on the equation
for $(b-a)$ yields the following wave equations for $c_{1}$
and $c_{2}$:
\be
\left( -\laptil + \frac{2}{R^{2}} \right) c_{1} = 0 , \; \; \; 
\left( -\laptil + \frac{6}{R^{2}} \right) c_{2} = 0 .
\label{c1c2}
\ee
Since the operators are positive, we may use the standard argument
to conclude that Eqs. (\ref{c1c2}) admit neither unstable nor 
well-behaved stationary modes. Hence, $c_{1} = c_{2} = 0$, 
implying that $a = b$ and $c=0$. It therefore remains to
solve $\diff (R^{2} \tilast \diff a) = 0$ for the gauge-invariant 
scalar field $\tilast \diff a$. With respect to
Schwarzschild coordinates, the result is $a = b = (q/r) \diff t$
plus gauge terms, where $q$ is a constant of integration.
Now using $\alpha = a + \diff \mu$, $\beta = b$ and
$\nu = c +\mu$ in Eq. (\ref{YMP-4}) gives
$\delta A^{(\ell = 1)} =a X_{1} + b X_{2} + c Y \diff \taur +
\diff (\mu X_{1})$, and thus, with $c=0$, $a = b = (q/r) \diff t$ and
$X_{1} + X_{2} = \tau_{z}$,
\bdm
\delta A = \tau_{z} \frac{q}{r} \diff t
\edm
plus a pure gauge term. Using
$\delta F = \Diff \delta A = \diff \delta A$ for $w=1$, this
gives rise to the electric field
\be
\delta E = -\tau_{z} \frac{q}{r^{2}} \diff r .
\label{Sch1-6}
\ee
The solutions (\ref{Sch1-1}) and (\ref{Sch1-6}) describe the 
Kerr-Newman excitation of the Schwarzschild metric in first order of
the rotation parameter $a$ and the electric charge $q$.

\subsection{Perturbations with $\ell = 0$}
\label{section-Schw-C}

The relevant perturbation equation is Eq. (\ref{S-2}) with
$\bar{B} = H = 0$, $w=1$ and $\lambda = -2$. The amplitude $A = \bar{A}$ 
is gauge-invariant and, by virtue of Eq. (\ref{S-12}), well-defined. 
Equation (\ref{S-2}) becomes
$\cod (R^{2} \diff A) + 2A = 0$. Using the integrability condition 
$\cod A = 0$, the scalar field $\Psi$ is defined according to 
$\Psi \equiv R^{2} \tilast \diff A$, in terms of which Eq. (\ref{S-2}) becomes
\be
\left( -\laptil + \frac{2}{R^{2}} \right) \Psi = 0,
\label{Sch0-0}
\ee
which admits neither unstable nor acceptable stationary solutions. 
(Note that the RN excitations with $\ell = 0$ of the 
Schwarzschild metric lie in the {\it even\/} parity sector.)

\section{Stationary perturbations of non-Abelian solitons and black holes}
\label{section-SPEYM}

Having analyzed the complete set of non-Abelian odd-parity perturbations 
(stationary and dynamical) of the Schwarzschild and the RN solutions, 
we now turn to the general case, that is, to non-Abelian perturbations of 
static non-Abelian background configurations. 
The discussion of the corresponding
perturbation equations is a considerably more involved task, since the 
techniques used above cannot be applied if $w$ is not constant.
Our primary goal in this section is to classify all {\it stationary\/} 
odd-parity perturbations of both the BK solitons \cite{Bartnik88} and the 
static, spherically symmetric EYM black holes \cite{VolkGal89}.

In the stationary case, the excitations of a spherically symmetric EYM 
background decouple into two Sturm-Liouville problems, governing the 
electric and the magnetic perturbations, respectively. The particular case
$\ell = 1$ was analyzed in Refs. \cite{BHPRD} and \cite{BHSV} by different 
means. There we have shown that the electric 
sector gives rise to a two parameter family of slowly rotating and / or
electrically charged black hole excitations, and to
a one-parameter family of slowly rotating, electrically charged solitons. 
In this section we 
generalize these results as follows: We show that for {\it all\/} values of 
$\ell \geq 1$ the electric perturbations are governed by a
three-channel Sturm-Liouville problem, while the
magnetic sector is described by a single Sturm-Liouville equation
for $\ell > 1$ and is trivial for $\ell = 1$. 
A careful analysis then reveals that 
neither the electric nor the magnetic sector admit well-defined stationary 
soliton or black hole excitations if $\ell > 1$. This establishes the result 
that the only stationary odd-parity modes of the BK solitons and EYM black 
holes are the ones found in Ref. \cite{BHSV} for $\ell = 1$.

It turns out to be convenient to 
parametrize the two-dimensional background 
metric $\gtiltens$ in terms of the radial coordinate $\rho$, defined such 
that $\gtiltens$ becomes conformally flat,
\be
\gtiltens = - N S^{2}  \diff t^{2} + \frac{1}{N} \diff r^{2} = 
\sigma \left( -\diff t^{2} + \diff \rho^{2} \right) ,
\label{SP-1}
\ee
with $\sigma(\rho) \equiv N (r) S^{2}(r)$ and 
$\diff r \equiv N S \diff \rho$. [The coordinate $\rho$
generalizes the coordinate $r_{\star}$ used in the Schwarzschild or
the RN case. We also note that
$\tilast \diff t = - \tilast \diff \rho$ and
$\sigma \tilast ( \diff t \we \diff \rho) = -1$.]
The invariant one-forms $A$, $B$ and $H$ are expanded with respect to 
$t$ and $\rho$, e.g.,
\be
H \equiv H_{0} \diff t + H_{1} \diff \rho .
\label{SP-2}
\ee
Since we restrict ourselves to stationary perturbations the coefficients 
$H_{0}$, $H_{1}$, etc. are functions of $\rho$ only. As we shall argue below,
the equations involving the zero-components, henceforth called electric 
perturbations, decouple from the equations for the one-components, henceforth 
called magnetic perturbations.

\subsection{The electric sector}
\label{subsection-SPEYM-1}

The electric perturbation equations involve the amplitudes $H_0$, $A_0$ 
and $B_0$ only. [For $l = 1$, we may take $H_0 = h_0$, $A_0 = a_0$ and 
$B_0 = b_0$, since, by virtue of Eqs. (\ref{GE-4}), (\ref{YMP-8}) and 
(\ref{Lie-4}), these amplitudes are invariant under {\it stationary\/} 
coordinate transformations.] Using the fact that 
$[\cod (R^2 \diff A)]_0 = -\p_\rho (\sigma^{-1} R^2\p_\rho A_0)$
for stationary perturbations of a static background, the zero-components 
of Eqs. (\ref{E2}), (\ref{delYM-1}) and (\ref{delYM-2}) may be cast into 
the following three-channel Sturm-Liouville equation:
\be
\left( -\p r^2 \p + \bbK\p - \p\bbK^T + \bbL + \bbP \right)\ul{v} = 0,
\label{ODD_ELEC}
\ee
where $\ul{v} \equiv \left(H_0/\sqrt{4G}r, A_0/\zeta, B_0\right)$, 
$\zeta \equiv \sqrt{\lambda+2}$, and where the differential operator 
$\p$ is defined by
\bdm
\p \equiv \frac{1}{\sigma}\frac{\diff}{\diff\rho} = 
\frac{1}{S}\frac{\diff}{\diff r},
\edm
with $r$ and $\rho$ according to Eq. (\ref{SP-1}).
The $3\times 3$ matrices $\bbK$, $\bbL$ and $\bbP$ are given in terms of 
the background fields $w$, $N$ and $\sigma = S^2 N$. The only 
non-vanishing matrix element of $\bbK$ is $\bbK_{13}=\sqrt{4G} r\,\p w$, 
while the symmetric matrices $\bbL$ and $\bbP$ are
\be
\bbL = \frac{1}{\sigma}\left( \begin{array}{ccc}
2N + \lambda &          sym.       & sym.  \\
           0 & \lambda+2(1+w^2) & sym. \\
           0 & -2\zeta w          & \lambda + (1+w^2)\\
\end{array} \right),
\nonumber
\ee
and
\be 
\bbP = \frac{1}{\sigma}\left( \begin{array}{ccc}
4G\frac{(w^2-1)^2}{r^2} + 2G\sigma (\p w)^2  & sym. & sym.  \\
\sqrt{4 G} \zeta\frac{1-w^2}{r} & 0 & sym. \\
\sqrt{4 G} w \frac{w^2-1}{r} & 0 & 0\\
\end{array} \right).
\nonumber
\ee
The formally self-adjoint equation (\ref{ODD_ELEC}) holds for all values
of $\ell\geq 1$. [In particular, for $\ell = 1$ it is 
equivalent to the Sturm-Liouville equation derived in \cite{BHPRD}, which 
was shown to admit the stationary modes mentioned above \cite{BHSV}.
However, the transformation between the two $\ell = 1$ sets of 
equations is not algebraic, because the original formulation given in
\cite{BHPRD} was based on the generalized twist potential.]

Since Eq. (\ref{ODD_ELEC}) has {\it regular singular\/} points at the origin, 
$r=0$, at the horizon, $r=r_H$ (where $N(r_H)=0$), and at infinity, $r=\infty$,
it is possible to compute the number of stationary modes.
Applying the standard theory (see, e.g., \cite{Walter}) we will now discuss
the local solution spaces.

\subsubsection{The solution space at the origin}

The leading order behavior of the solutions to Eq. (\ref{ODD_ELEC}) 
in the vicinity of the origin is determined by 
the centrifugal barrier $\bbL$, as can be seen from the
expansions (\ref{EYMEXP1}) of the background quantities. 
The solutions behave like $r^\alpha$, where $\alpha=-(\ell+2)$, 
$-(\ell+1)$, $-\ell$, $\ell-1$, $\ell$ or $\ell+1$. 
Hence, the space of regular solutions at $r=0$ is {\it three}-dimensional 
for all values of $\ell\geq 1$. The expansion becomes
\bea
\ul{v}(r) &=& d_1 r^{\ell-1} \! \!\left[ \ul{e}_- \!\! + 
              \frac{\ell \!-\! 1}{2 \ell + 1}
       \left( 2(\ell \! + \! 2)b+1\right)b r^2 \ul{e}_- \! + {\cal O}(r^3) \right]
\nonumber\\ &+&
              d_2 r^{\ell} \left[ \ul{e}_0 + 
        \frac{2 b}{2 \ell \! + \! 1}
       r \ul{e}_- \! + {\cal O}(r^2) \right]
\nonumber\\ &+&
            d_3 r^{\ell+1} \left[ \ul{e}_+ + {\cal O}(r) \right]
\label{LS0}
\eea
where $\ul{e}_0 = (1,0,0)$, $\ul{e}_+ = (0,\zeta,-l)$ and 
$\ul{e}_- = (0,\zeta,l+1)$, and $d_1$, $d_2$ and $d_3$ are constants,
and where $b$ is the fixed constant appearing in the expansions 
(\ref{EYMEXP1}) of the background solutions.

\subsubsection{The solution space at infinity}

The asymptotic expansions (\ref{EYMEXP2}) of the background quantities show 
that the leading order behavior of the solutions to Eq. (\ref{ODD_ELEC}) 
is again completely determined by $\bbL$: 
The solutions behave like $r^\alpha$,
where again 
$\alpha=-(\ell+2)$, $-(\ell+1)$, $-\ell$, $\ell-1$, $\ell$ or $\ell+1$. 
The space of asymptotically flat solutions is, therefore, 
{\it three}-dimensional for $\ell>1$, and {\it four}-dimensional for $\ell=1$.
For $\ell=1$ the asymptotic expansion is found to be
\bea
\ul{v}(r) &=& \left( c_0 + \frac{c_1}{r} \right) \! 
\left[ \ul{e}_- \! + {\cal O}\!\left(\frac{\log r}{r^2}\right) \right] 
+ \frac{c_2}{r^2}\left[ \ul{e}_0 + \! 
{\cal O}\left(\frac{1}{r^2}\right) \right] 
\nonumber\\
          &+& \frac{c_3}{r^3}\left[ \left(1+(1-\gamma)\frac{2M}{r}\right)
\ul{e}_+ + \! {\cal O}\left(\frac{1}{r^2}\right) \right] .
\label{LSinf}
\eea

The constant $c_2$ is proportional to the total angular momentum $\delta J$, 
while $c_0$ and $c_1$ are proportional to the asymptotic value of the 
electric YM potential $\delta \Phi_\infty$ and the electric YM charge 
$\delta Q_e$, 
respectively: Using the above expansion in the expressions 
(\ref{FLUX}) for the linearized local Komar integrals, we find
[with $\tilast F_B = \sigma^{-1} (B_0'+w'H_0/R^2)$ etc.],
\bdm
\delta Q_e(r\rightarrow\infty) \sim \ul{e}_m\cdot\ul{\tau}\, c_1, \;\;\;
\delta J(r\rightarrow\infty) \sim \delta_{m\,0}\, c_2.
\edm
Furthermore, the above expansion, together with the definition
(\ref{l=1g-and-A}) and $\delta\Phi = \delta A(\partial_t)$, 
shows that $\delta\Phi_\infty$ is proportional to $c_0$.
It is worthwhile recalling that, in contrast to the Abelian case, 
$c_0$ cannot be ``gauged away''. This is also obvious form the fact that
the expression for $\delta F$ involves an asymptotically vanishing term
proportional to $c_0/r$, unless for $w=1$.

\subsubsection{The solution space at the horizon}

Using the background expansions (\ref{EYMEXP3}) at the horizon,
the solutions to Eq. (\ref{ODD_ELEC}) behave like $(r-r_H)^\alpha$,
where the eigenvalues are $\alpha=0$ and $\alpha=1$, and the multiplicity
is three in both cases. For $\alpha=0$ the three eigenvectors may pick up 
logarithmic terms in next-to-leading order, which destroy the 
regularity of the horizon. A careful analysis shows that the number of 
eigenvectors with logarithmic terms in next-to-leading order is equal to the 
rank of the symmetric matrix
\bdm
\bbS_1 = \left( \begin{array}{ccc}
\lambda + 4G\frac{(w_H^2-1)^2}{r_H^2} & sym.                & sym. \\
\sqrt{4G}\zeta\frac{1-w_H^2}{r_H}       & \lambda+2(1+w_H^2)  & sym. \\
\sqrt{4G} w_H\frac{w_H^2-1}{r_H}      & -2\zeta w_H & \lambda + 1+w_H^2
\end{array} \right) ,
\edm
which is proportional to the leading order term of $\bbL + \bbP$ 
in $r-r_H$. The determinant of $\bbS_1$ is given by
\bdm
\det\bbS_1=\lambda\left[ \lambda^2 + (3-w_H^2)\lambda + 2(1-w_H^2)^2 + 8G\,G_H^2 \right],
\edm
where we recall that $w_H \equiv w(r_H)$ and $G_H \equiv w_H (w_H^2-1)/r_H$.
This shows that the rank of $\bbS_1$ is three for $\ell>1$, while one
may also verify that the rank is two for $\ell=1$. Hence, all solutions with 
$\alpha = 0$ must be excluded, unless $\ell = 1$, in which case there exists
one acceptable eigenvector.
The physical space of solutions at $r=r_H$ is, therefore, 
{\it three}-dimensional for $\ell>1$ and {\it four}-dimensional for $\ell=1$.

\subsubsection{Soliton excitations}
 
Since the BK background is continuous, and since the perturbation equations 
are linear with continuous coefficients for $0 < r < \infty$, 
the local solutions 
(\ref{LS0}) and (\ref{LSinf}) admit extensions to the semi-open intervals 
$[0,\infty)$ and $(0,\infty]$, respectively. Since, for $\ell = 1$, these 
solution subspaces are three- and four-dimensional, respectively, and since
the total space of solutions is {\it six}-dimensional, we conclude that 
the intersection space is generically one-dimensional. Hence, there exists
(at least) one global solution, describing the rotating charged solitons 
found in \cite{BHSV}.

For $\ell>1$. the intersection space is generically trivial, since the
solution spaces are three-dimensional at both the origin {\it and\/}
infinity. Hence, there exist no generic soliton excitations for $\ell > 1$.
In fact, {\it non-generic} solutions are excluded as well, 
as we shall prove below.

\subsubsection{Black hole excitations}

Applying the same argument as in the soliton case, we conclude that 
Eq. (\ref{ODD_ELEC}) admits a {\it two}-dimensional intersection space of global 
solutions for $\ell = 1$, since the local solution spaces at the
horizon and at infinity are four-dimensional. The solutions give rise
to the black hole excitations found in \cite{BHSV}, which are parametrized 
by their total angular momentum $\delta J$ an their electric YM charge 
$\delta Q_e$.

For $\ell>1$ there exist again no generic solutions, since the
solution spaces at the horizon and at infinity are three-dimensional only.
It therefore remains to exclude non-generic solutions, 
which we shall do next.

\subsubsection{Absence of non-generic solutions for $\ell>1$}

Our aim is to show that Eq. (\ref{ODD_ELEC}) with the boundary 
conditions discussed above admits neither soliton nor black hole 
solutions for $\ell > 1$.
We do so by casting Eq. (\ref{ODD_ELEC}) into the form required to 
apply the argument outlined in Appendix \ref{App-B}. This is achieved 
by performing the linear transformation $\ul{v}=\bbT\ul{u}$, which yields
\be
\left( -\p \bbA \p + \bbS \right)\ul{u} = 0 ,
\label{SYMPOS}
\ee
where $\bbA$ is symmetric and positive, while $\bbS$ is symmetric and 
positive semidefinite. The linear transformation $\bbT$ is given by 
$\bbT=\bbT_1\circ\bbT_2$, where
\bdm
\bbT_1 = \mbox{diag}(r,1,1), \;\;\;
\bbT_2 = \idid - \sqrt{4G} \left( \begin{array}{ccc}
 0 & 0 & 0 \\
 1/\zeta & 0 & 0 \\
 w & 0 & 0 \\
\end{array} \right)
\edm
[Note that the components of $\ul{u} = \bbT^{-1}\ul{v}$ coincide
with the amplitudes introduced in Eq. (\ref{bars}):
$\sqrt{4G} u_1 = H_0/r^2$, $\zeta u_2 = A_0 + H_0/r^2$,
$u_3 = B_0 + w H_0/r^2$.] The Sturm-Liouville equation (\ref{ODD_ELEC})
now assumes the desired form (\ref{SYMPOS}), with the symmetric matrices
$\bbA=r^2 \bbT^T\circ\bbT$, 
$\bbS=\bbT_2^T\circ\tilde{\bbS}\circ\bbT_2\,$, where
\bdm
\tilde{\bbS} = \frac{1}{\sigma}\left( \begin{array}{ccc}
\lambda r^2 \!+\! 4G(w^2\!-\!1)^2 & sym.                & sym.            \\
\sqrt{4G}\zeta (1-w^2)      & \lambda \!+\!2(1\!+\!w^2)      & sym.            \\ 
\sqrt{4G} w(w^2-1)        & -2\zeta w     &  \lambda \! + \!1\!+\!w^2
\end{array} \right).
\edm
It is not hard to see that the matrix $\tilde{\bbS}$ is positive 
for all values of $\ell > 1$ and positive semidefinite for $\ell=1$. 
Furthermore, by virtue of the expansions given above for $\ell > 1$, 
the boundary term $\ul{u}\cdot\bbA\p\ul{u}$ vanishes at the origin, 
at the horizon, and at infinity. Both soliton and black hole solutions
are, therefore, excluded as a consequence of the argument given 
in Appendix \ref{App-B}.

We emphasize that the boundary terms at the origin and at the horizon
do give non-vanishing contributions if $\ell = 1$. The positive
operator in Eq. (\ref{SYMPOS}) is, therefore, self-adjoint
only for $\ell > 1$.

\subsubsection{Conclusion}

We have proven the following {\it local uniqueness theorems\/}
for odd-parity perturbations in the electric sector: The only stationary, 
asymptotically flat {\it black hole\/} solutions which are infinitesimally 
close to the static, spherically symmetric EYM black holes 
are the rotating and/or electrically charged excitations in the $\ell=1$ sector.
The only {\it soliton\/} solutions which are infinitesimally close to the 
BK solitons are the electrically charged excitations in the $\ell=1$ sector.

These results are in agreement with the non-Abelian staticity 
theorem \cite{SudWald92}, which asserts that spacetime is static and 
purely magnetic if the combination 
$\Omega_H J - \Trace{\Phi_\infty Q_e}$ vanishes, where
$\Omega_H$ is the angular velocity of the horizon: 
For $l=1$, non-static solitons and black holes can exist, while, 
for $\ell > 1$, there is no contribution to $J$ and $Q_e$ 
[see the general formulae (\ref{FLUX})], implying that the 
non-static and electric contributions 
$H_0$, $A_0$ and $B_0$ must vanish.

\subsection{The magnetic sector}
\label{subsection-SPEYM-2}

For stationary perturbations one has $\cod A = -\sigma^{-1} A_{1}'$, 
where here and in the following a prime denotes differentiation with 
respect to the radial coordinate $\rho$, defined in Eq. (\ref{SP-1}).
Since the background is static, one also has 
$\sprod{A}{\diff w} = -\sigma^{-1} w' A_{1}$. Hence, the gravitational 
constraint (\ref{E1}) and the YM constraints (\ref{YMint-1}), 
(\ref{YMint-2}), as well as Eq. (\ref{delYM-3}) involve only the 
one-components of $A$, $B$ and $H$. It is, therefore, possible to express
the YM amplitudes $A_{1}$, $B_{1}$ and $C$ in terms of the gravitational 
perturbation $H_{1}$:
\bea
A_{1} & = & \left(w^{2}-1 + 
\frac{R^{2}}{2 G} \right) \frac{H_{1}}{R^{2}} , 
\nonumber\\
B_{1} & = & \frac{1}{4 G w'}  H_{1}' ,
\nonumber\\
C & = & - \frac{1}{\sigma} \left[
w' H_{1} + \frac{R^{2}}{4 G} \left( \frac{H_{1}'}{w'} \right)'
\right].
\label{SP-3}
\eea
Using the above expressions and the circumstance that
$[\cod (R^{4} \diff(H/R^{2}))]_{1}$ vanishes for stationary perturbations 
of a static background, the one-component of the gravitational 
equation (\ref{E2}) yields the following Sturm-Liouville equation for $H_{1}$:
\be
\left[ -\frac{\diff}{\diff\rho} \frac{1}{w'^2} \frac{\diff}{\diff\rho} +
\frac{[\ell (\ell \!+\! 1)\! -\! 2 w^2] \sigma \!-\! 4 G w'^2}{R^2 w'^2}
\right] H_{1} = 0 ,
\label{SP-4}
\ee
where we recall that 
$w' = \diff w /\diff \rho = N S \diff w / \diff r$.

The above equation holds for $\ell > 1$ only. For $\ell = 1$ the perturbations 
are governed by Eqs. (\ref{EYM-03bis}). 
Since the one-components of the first terms in these equations vanish 
for stationary perturbations, we obtain $\bar{a}_{1} = \bar{b}_{1} = 0$, 
provided that $w^2 - 1$ does not vanish everywhere. Now using the
fact that there exists a gauge for which $h_{\rho}$ vanishes if 
$\ell = 1$, we conclude that magnetic excitations cannot exist for $\ell = 1$.
(The case $w^2=1$, $\ell = 1$ has already been discussed in 
Sect. \ref{section-Schw-B}.)

Equation (\ref{SP-4}) has regular singular points at the origin, $R = 0$, 
at the horizon, $N = \sprod{\diff R}{\diff R} = 0$, at infinity, $R = \infty$, 
and at all points where $w'$ vanishes. (For the one-node background 
solutions this is only the case at the origin and at infinity.) In order 
to conclude that Eq. (\ref{SP-4}) generically admits neither acceptable 
soliton nor black hole excitations, it is sufficient to discuss the regular 
singular points at the boundaries in leading order.

\subsubsection{Soliton excitations}

Using the expansions (\ref{EYMEXP1}) for the BK background at the origin 
shows that the fundamental solutions to Eq. (\ref{SP-4}) behave like
$r^{\ell + 2}$ and $r^{1- \ell}$. Since $\ell > 1$, the subspace of solutions
giving rise to finite metric perturbations is, therefore, 
{\it one}-dimensional at the origin. 
In the asymptotic region one uses the expansions (\ref{EYMEXP2}) to
conclude that the fundamental solutions behave like $r^{-\ell - 2}$ 
and $r^{\ell-1}$, implying that the subspace of bounded solutions is 
again {\it one}-dimensional. Generic soliton excitations are, therefore, 
excluded.
[The subspace of bounded solutions at the inner points $w'=0$
turn out to be two-dimensional. It is, however, generically not possible to
match the solutions from $r=0$ and $r=\infty$ at the points $w'=0$ 
such that the amplitude $C$ is continuous.]

\subsubsection{Black hole excitations}

Using the horizon expansions (\ref{EYMEXP3}) shows that the fundamental 
solutions to Eq. (\ref{SP-4}) behave like $(r-r_H)^0$ and $(r-r_H)^2$.
The first solution is physically unacceptable, since the
invariant quantity $\sprod{H}{H} = H_1^2/\sigma$ diverges for
$r \rightarrow r_H$. Hence, the physical subspaces at the horizon
and at infinity are {\it one}-dimensional, implying that black hole 
excitations do not exist in the generic case. 

So far, we were not able to exclude non-generic solutions by rigorous 
means: The first problem is that the potential in Eq. (\ref{SP-4}) is 
not manifestly positive (although numerical investigation suggest that 
this is the case). Furthermore, the boundary term arising in the integral 
argument given in Appendix \ref{App-B} does not vanish at points 
where $w'=0$. It is, however, clear that the potential is positive if
$\ell$ is big enough. In this case the integral argument applies, 
at least for excitations of the background solutions with one node.

\subsubsection{Conclusion}

Since $H_1$ parametrizes the {\it non-circular\/} part of the metric, 
we have shown that there exists no non-circular deformations in the 
odd-parity sector. This completes the classification of the stationary
odd-parity excitations of the BK solitons and the corresponding
non-Abelian black holes.
The only physically admissible non-Abelian stationary odd-parity
excitations of these configurations are the rotating, electrically 
charged solitons and the two-parameter family of black holes found in
\cite{BHSV}. All modes lie in the electric part of the
distinguished sector $\ell = 1$.

\section{Dynamical perturbations}
\label{sect-nonstationary}

Stationary perturbations need to be analyzed in order to find
equilibrium solutions which are infinitesimally neighbored to
known static configurations, or to establish local uniqueness
results. The linear {\it stability\/} properties of static 
background solutions are, however, described by 
{\it non-stationary\/} perturbations. 
In order to study their dynamical behavior by means of 
spectral theory, it is necessary to cast the perturbation 
equations into a system of {\it pulsation\/} equations, that is, 
into a wave equation whose spatial part is (formally) {\it self-adjoint}.
Using the static EYM soliton or black hole background, our task is,
therefore, to write the perturbation 
equations (\ref{S-1}) - (\ref{S-4a}) in the form
\be
\left[ \frac{\p^2}{\p t^2} + \bbcA \right] u = 0,
\label{PULS}
\ee
where $\bbcA$ is a self-adjoint operator, containing spatial derivatives up
to second order. For perturbations of the Schwarzschild and
RN black holes this was achieved in Eqs. 
(\ref{Psi12-eq2}), (\ref{Pi-i}), (\ref{Schw1}) and
(\ref{Schw3}).
For perturbations of non-Abelian background configurations, however, 
one needs to proceed differently:

For $\ell = 0$ (i.e., for radial perturbations), the above task was achieved 
in \cite{Volkov}, where it was shown that the static, spherically symmetric
BK solitons and EYM black holes have exactly $n$ unstable radial modes in the 
odd-parity sector, $n$ being the number of nodes of $w$.

For $\ell = 1$, we will show below that the metric perturbations decouple, and 
that the perturbation equations can be cast into a wave equation for the 
remaining YM perturbations, where the operator ${\cal A}$ is symmetric and 
positive. This will establish the absence of unstable
odd-parity modes in the sector $\ell = 1$.

For $\ell > 1$, we were not able to derive symmetric equations in terms of 
the gauge-invariant amplitudes $H$, $A$, $B$ and $C$. However, a system of
hyperbolic equations can be obtained as follows: By virtue of 
Eqs. (\ref{E1}), (\ref{delYM-3}) and (\ref{YMint-1}) one can express the 
time derivatives of the electric components $H_0$, $A_0$ and $B_0$ in terms 
of the magnetic components $H_1$, $A_1$, $B_1$ and $C$ and their first
spacial derivatives. Equations (\ref{E2}), 
(\ref{delYM-1}), (\ref{delYM-2}) and (\ref{YMint-2}) then yield a hyperbolic 
system of the form
\bdm
\left[ \frac{\p^2}{\p t^2} - \frac{\p^2}{\p\rho^2} +  \bbK\frac{\p}{\p\rho} + \bbV \right] u = 0,
\edm
where $u$ comprises the magnetic components,  $u = (H_1, A_1, B_1, C)$, 
and where the radial coordinate $\rho$ is defined as 
in (\ref{SP-1}). Unfortunately, neither the first order derivatives 
nor the potential $\bbV$ are formally self-adjoint.

In \cite{OddLetter} we have argued that the gauge-invariant amplitudes 
used in the present paper are not suited to describe dynamical 
perturbations, an exception being vacuum gravity or self-gravitating
Abelian fields. 
In order to obtain a symmetric wave equation one needs to introduce 
amplitudes which are adapted to the {\it staticity\/} rather than the 
spherically symmetry of the background. In terms of these 
new, {\it curvature-based\/} amplitudes, the odd-parity pulsation 
equations can be cast into the desired form (\ref{PULS}), as we
have shown in \cite{OddLetter}.

In the remainder of this section we present the distinguished
cases $\ell = 0$ and $\ell = 1$.
In these situations the perturbation equations can be
written in the desired form, since gravity can be decoupled for $\ell = 1$,
while for $\ell = 0$ only YM perturbations are present 
in the odd-parity sector.
 
\subsection{The pulsation equation for $\ell = 0$}
\label{subsectionstationary-0}

For spherically symmetric perturbations the pulsation equation is obtained 
from Eq. (\ref{EYM-02bisbis}),
\bdm
\cod \left(R^{2} \diff a \right) + 2 w^{2} a = 0 ,
\edm
for the gauge-invariant YM amplitude $a = \alpha - \diff ( \nu / w ) $; 
see Eq. (\ref{1-ell-0}). This equation is not regular at points where 
$w$ vanishes. (The case where $w$ vanishes identically was discussed in 
Sect. \ref{section-RN-3}. Introducing the regular one-form 
$w^2 a = w^2 \alpha + \nu \diff w - w \diff\nu$, and defining the potential 
$\Phi$ by the equation $w^2 a = \tilast\diff (w\Phi)$, we find
\bdm
-\laptil\Phi + \left[ 2 \sprod{\frac{\diff w}{w}}{\frac{\diff w}{w}}
+ \frac{1}{R^2}(w^2+1) \right] \Phi = 0,
\edm
where we have also used the background YM equation (\ref{YMBG-4}).
For a static background we may assume a time dependence of the form 
$\exp (i\omega t)$, which yields
\be
\left[ -\frac{\p^2}{\p\rho^2} + 2\frac{w'^2}{w^2} + \frac{\sigma}{R^2}(w^2 + 1) \right] \Phi = \omega^2 \Phi,
\label{ODDl=0}
\ee
where a prime denotes the differentiation with respect to $\rho$, and
now $\Phi \equiv \Phi(\rho)$. In order to overcome the difficulty that the
potential is singular at points where $w$ vanishes, one may perform the 
following super-symmetric transformation: First, the operator on the LHS 
can be factorized and written as $Q^{\dagger} Q$, with $Q$ and $Q^{\dagger}$ 
according to
\bdm
Q = \frac{1}{w} \frac{\p}{\p\rho} w + u, \;\;\;
Q^{\dagger} = -w \frac{\p}{\p\rho} \frac{1}{w} + u ,
\edm
where $u$ is subject to the differential equation
\be
-w^2 \left( \frac{u}{w^2} \right)' + u^2 = \frac{2\sigma w^2}{R^2}.
\label{AUX}
\ee
One may then write Eq. (\ref{ODDl=0}), 
$Q^{\dagger} Q \Phi = \omega^2 \Phi$, in terms of $\Psi \equiv Q\Phi$, which
yields $Q Q^{\dagger} \Psi = \omega^2 \Psi$. Since 
$\omega^2\Phi = Q^{\dagger} \Psi$, there is a one-to-one correspondence 
between $\Phi$ and
$\Psi$, provided that $\omega \neq 0$. Furthermore, 
$\Psi$ is normalizable if $\Phi$ is normalizable, and vice-versa, since 
$\sprod{\Psi}{\Psi} = \sprod{Q\Phi}{Q\Phi} = \sprod{Q^{\dagger} Q \Phi}{\Phi} 
= \omega^2 \sprod{\Phi}{\Phi}$.

The equivalent problem, $Q Q^{\dagger} \Psi = \omega^2 \Psi$, reads
\be
\left[ -\frac{\p^2}{\p\rho^2} + \frac{\sigma}{R^2}(3w^2 - 1) + 
2u' \right] \Psi = \omega^2 \Psi,
\label{ODDTransl=0}
\ee
where now the potential is regular, provided that $u$ is a regular
solution of Eq. (\ref{AUX}). Since the function
\bdm
\Psi_0 = w \exp\int_{\rho_0}^{\rho} u(\tilde{\rho})\, \diff\tilde{\rho}
\edm
satisfies $Q^{\dagger}\Psi_0 = 0$, it is a solution to Eq. (\ref{ODDTransl=0})
for $\omega = 0$.
The key observation in \cite{Volkov} is that there exists a solution to
Eq. (\ref{AUX}) such that $u/w^2$ and $u'$ are regular 
and $\Psi_0$ is normalizable.
Since the factor $w$ causes $\Psi_0$ to have exactly $n$ nodes 
($n$ being the number of nodes of $w$), this establishes the fact that
the transformed pulsation equation (\ref{ODDTransl=0}) admits
exactly $n$ unstable modes.

It remains to show that each unstable mode of Eq. (\ref{ODDTransl=0}) 
can be realized by a regular choice of the original amplitudes 
$\alpha$ and $\nu$. In order to see this, we first note that for 
$\omega\neq 0$ the inverse transformation becomes
\bdm
w\Phi = \frac{1}{\omega^2} \left( -w \Psi' + w'\Psi + w u \Psi \right),
\edm
implying that the gauge-invariant combination $w^2 a$ is regular.
Finally, one adopts the {\it temporal gauge\/}, $\alpha_t = 0$,
with respect to which Eq. (\ref{ODDl=0}) yields
\bdm
\frac{\p}{\p t} \alpha_\rho = \frac{2\sigma}{R^2} w\Phi,
\edm
implying that $\alpha_\rho$ is regular. Using 
$\Psi = Q\Phi = w^{-1} (w\Phi)' + u\Phi$, as well as the $t$-component of 
$w^2 a = \tilast\diff (w\Phi)$ in the temporal gauge, gives
\bdm
\frac{\p}{\p t} \nu = \Psi - \frac{u}{w} w\Phi.
\edm
This establishes the existence of exactly $n$ unstable modes of the 
original perturbation equations, since $u/w$ can be chosen to be regular,
implying that $\nu$ is regular.

\subsection{The pulsation equation for $\ell = 1$}

We now show that for $\ell = 1$ the gravitational perturbations can be
expressed in terms of the YM perturbations, which yields a pulsation 
equation for the YM amplitudes. The gravitational amplitude $h$ enters 
the perturbation equations (\ref{EYM-03bis})
only via thecoordinate-invariant combination $F_h=\diff( R^{-2} h)$.
The crucial observation is that the second plus $2w$ times the third minus 
$2G$ times the first equation in (\ref{EYM-03bis}) yields
the conservation law
\bdm
\cod \left[ -\frac{1}{2G} R^4 F_h + R^2 \left( F_a + 2w F_b \right) \right] = 0.
\edm
Recalling the definitions $F_a = \diff\bar{a} - F_h$ and  
$F_b = \diff\bar{b} - w F_h$, we find after integrating the above equation
\bdm
F_h = f \left( \diff\bar{a} + 2w\, \diff\bar{b} + \frac{c_0}{R^2}\tilast 1 \right),
\edm
where $c_0$ is a constant, and $f$ denotes the background quantity
$f \equiv (R^2/2G + 1 + 2w^2)^{-1}$.
Using this expression for $F_h$ in Eqs. (\ref{EYM-03bis}) yields 
the symmetric, inhomogeneous equation
\be
\cod \left[ \bbG\,\diff \left( \begin{array}{c} 
\bar{a} \\ \bar{b} \end{array} \right) \right] + 
\bbF        \left( \begin{array}{c} \bar{a} \\ \bar{b} \end{array} \right)
 = - c_0 \tilast\diff \left( \begin{array}{c} f \\ 2w f \end{array} \right)
\label{ODDl=1}
\ee
for the gauge- and coordinate-invariant YM amplitudes $\bar{a}$ and $\bar{b}$.
The $2\times 2$ matrices $\bbG$ and $\bbF$ are symmetric and given in terms
of the background quantities by
\bea
\bbG &=& \frac{f R^2}{2 } \left( \begin{array}{cc}
 4 w^2 + R^2 / G & -4  w     \\
 -4  w        & 4 + 2R^2 / G \\
\end{array} \right), \nonumber\\
\bbF &=& 2\left( \begin{array}{cc}
 1 + w^2 & -2w     \\
 -2w     & 1 + w^2 \\
\end{array} \right) , \nonumber
\eea
where $\bbF$ is positive definite for $w^2 \neq 1$. (The case $w \equiv 1$
was already discussed in Sect. \ref{section-Schw-B}.)

The one-forms $\bar{a}$ and $\bar{b}$ may be expanded with respect
to Schwarzschild coordinates $t$ and $\rho$,
\bdm
\left( \begin{array}{c} \bar{a} \\ \bar{b} \end{array} \right) =
E\,\diff t + B\,\diff\rho,
\edm
where $E$ and $B$ represent the gauge-invariant electric and magnetic YM 
fields, respectively. Using this in Eq. (\ref{ODDl=1}) gives, 
for a static background,
\bea
-\frac{\p}{\p\rho} \left[ \frac{1}{\sigma} \bbG 
\left( E' - \dot{B} \right) \right] + \bbF E 
&=& c_0 \frac{\p}{\p\rho} \left( \begin{array}{c} f \\ 
2w f \end{array} \right), \nonumber\\
-\frac{\p}{\p t}   \left[ \frac{1}{\sigma} 
\bbG \left( E' - \dot{B} \right) \right] + \bbF B 
&=& 0, \nonumber
\eea
where the dot and the prime denote differentiations with respect to 
$t$ and $\rho$, respectively. In particular, 
for stationary perturbations, $\dot{E} = \dot{B} = 0$, 
we recover the facts that the electric and the magnetic 
perturbations decouple, and that $B$ vanishes.

For dynamical perturbations a homogeneous pulsation equation of the desired 
form is obtained as follows: Differentiating the first equation with 
respect to $t$ and the second one with respect to $\rho$ yields
the relation
\bdm
\bbF\dot{E} = (\bbF B)' ,
\edm
where we have also taken advantage of the fact that the background is static.
Using this to eliminate $\dot{E}$ from the second equation, we obtain the
following two-channel wave equation with formally self-adjoint spatial part:
\bea
\Big[&& \frac{\p^2}{\p t^2} - 
\bbQ \frac{\p}{\p\rho} \bbQ^{-2} \frac{\p}{\p\rho} \bbQ
+ \frac{\sigma}{R^2}
\left( \begin{array}{cc}
 w^2 + 2 & -3w     \\
 -3w       & 2w^2 + 1 
\end{array} \right)
\nonumber\\
&+& 
\frac{4 G}{R^4} \sigma (1 - w^2)^2 
\left( \begin{array}{cc}
1 & 0     \\
0 & 0 
\end{array} \right) \Big] \bbQ B = 0,
\eea
where $\bbQ$ satisfies $\bbF = 2 \bbQ^2$, 
\bdm
\bbQ = \left( \begin{array}{cc}
1 & -w \\
-w & 1 
\end{array} \right).
\edm

Since the operator is symmetric and positive, we conclude that the
spherically symmetric EYM solitons and black holes have no
unstable odd-parity excitations in the sector $\ell = 1$.

\appendix
\section{Linearized Ricci and Einstein tensors}
\label{App-A}

In this Appendix we give the expressions for the linearized
Christoffel symbols and the Ricci and Einstein tensors.
As we have argued in Sect. \ref{subsection-GE-CI2}, it is 
sufficient to compute the perturbations in the ODG. The 
Christoffel symbols for an arbitrary (not necessarily static) 
spherically symmetric spacetime are
\bea
&&\chris{A}{bc} = 0, \;\;\;  \chris{a}{Bc} = 0, \nonumber\\
&&\chris{A}{BC} = \chrishat{A}{BC} , \;\;\; 
\chris{a}{bc} = \christil{a}{bc} , \nonumber\\
&&\chris{A}{Bc} = \delta^{A}_{\; B} R^{-1} \nabtil_{c} R, \;\;\; 
\chris{a}{BC} = - \ghat_{BC} R \, \nabtil^{a} R, \nonumber
\eea
where $\nabtil$ denotes the covariant derivative operator
with respect to the two-dimensional metric $\gtil$ defined in 
Eq. (\ref{GE-1a}). In the ODG the metric perturbations 
(\ref{GE-2}) and their inverse become
\bea
&&\delta g_{ab} = \delta g^{ab} = 
\delta g_{AB} = \delta g^{AB} = 0 \, , \nonumber\\
&&\delta g_{Ab} = h_{b} S_{A} \, , \;\;\; 
\delta g^{Ab} = -  h^{b} S^{A} \, , \nonumber 
\eea
where all indices are raised with the background metric, 
i.e., $h^{b} \equiv \gtil^{ab} h_{a}$ and
$S^{A} \equiv g^{AB} S_{B} \equiv R^{-2} \ghat^{AB} S_{B}$. 
Using this, and the background metric (\ref{GE-1a}),
the perturbed Christoffel symbols,
$\delta \chris{\mu}{\alpha \beta} = 
\frac{1}{2} g^{\mu \nu} (\delta g_{\alpha \nu ; \beta} +
\delta g_{\beta \nu ; \alpha} - \delta g_{\alpha \beta ; \nu})$,
become in the ODG
\bea
&&\delta \chris{a}{bc}  = 0 , \;\;\; 
\delta \chris{A}{BC} = S^{A} \ghat_{BC} \, R \,  h^{a} \nabtil_{a} R,
\nonumber\\
&&\delta \chris{a}{BC}  = h^{a} \nabhat_{\{B} S_{C\}}, \;\;\; 
\delta \chris{A}{bc}  = S^{A} \nabtil_{\{b} h_{c\}}, \nonumber
\eea
and
\bea
&& \delta \chris{a}{Bc}  = S_{B} \gtil^{ad}
\left( \nabtil_{[c} h_{d]} - h_{d} R^{-1} \nabtil_{c} R \right),
\nonumber\\
&& \delta \chris{A}{Bc}  = h_{c} R^{-2} \ghat^{AD}
\nabhat_{[B} S_{D]}, \nonumber
\eea
where $\nabtil_{[b} h_{a]} \equiv 
\frac{1}{2} (\nabtil_{b} h_{a} - \nabtil_{a} h_{b})$ and
$\nabtil_{\{b} h_{a\}} \equiv 
\frac{1}{2} (\nabtil_{b} h_{a} + \nabtil_{a} h_{b})$.
It is now a straightforward task to compute the perturbed 
Ricci tensor in the ODG. Using
$\delta R_{\alpha \beta} = 
\delta \Gamma^{\mu}_{\, \alpha \beta ; \mu} - 
\delta \Gamma^{\mu}_{\, \alpha \mu ; \beta}$, 
one finds
\be
\delta R_{ab} = 0 \, , \;\;\;
\delta R_{AB} = \nabtil^{a} h_{a} \nabhat_{\{A} S_{B\}} \, ,
\label{pert-RabRAB}
\ee
and
\bea
\delta R_{Ab} & = & \frac{S_{A}}{R^{2}}
\nabtil^{a} \left[R^{4} \nabtil_{[b} \left(
h_{a]} R^{-2} \right) \right] +
\frac{h_{b}}{R^2} \ghat^{BC} \, \nabhat_{C}
\nabhat_{[A} S_{B]} 
\nonumber\\ & - & 
\frac{S_{A}}{R^{2}}
\left( R \laptil R + \sprod{\diff R}{\diff R} \right) h_b .
\label{pert-RAb}
\eea
In order to simplify the expression for $\delta R_{Ab}$
we take advantage of the background equations (\ref{GE-1b}) 
to write $R \laptil R + \sprod{\diff R}{\diff R} = 
1 - \frac{1}{2}\ghat^{AB}R_{AB}$. Also using
the transversality of the spherical vector harmonics,
$\ghat^{AB}\nabhat_{A}S_{B} = 0$, we find 
\bdm
2 \ghat^{BC} \nabhat_{C} \nabhat_{[A} S_{B]} =
-\left( \laphat S \right)_{A} = \ell (\ell + 1) S_{A} ,
\edm
which we apply in the second term of Eq. (\ref{pert-RAb}). 
Finally using the fact that $\delta G_{Ab} = 
\delta R_{Ab} - \frac{1}{2}S_{A}h_{b} g^{\mu \nu} R_{\mu \nu}$ 
in the ODG, we obtain (with the identity 
$R^{2} \gtil^{ab}R_{ab} = \ghat^{AB}G_{AB} = R^2 G^B_{\, B}$)
the result
\be
\delta G_{Ab} = \frac{S_{A}}{R^{2}} \left\{
\nabtil^{a} \!
\left[ R^{4} \nabtil_{[b} \left(
h_{a]} R^{-2} \right) \right] \!+\! \frac{h_b}{2} \left(
\lambda + R^2 G^B_{\, B} \right) \right\},
\label{del-GAb}
\ee
which holds in the ODG.
The expression for $\delta G_{AB}$ follows from the fact that
in the ODG $\delta G_{AB} = \delta R_{AB}$,
\be
\delta G_{AB} = 
\nabhat_{\{A} S_{B\}} \nabtil^{b} h_{b} .
\label{del-GAB}
\ee
Eventually,
$\delta G_{ab} = \delta R_{ab} - \frac{1}{2} R^{-2} \gtil_{ab}
\ghat^{AB} \delta R_{AB}$ in the ODG, which vanishes
by virtue of Eqs. (\ref{pert-RabRAB}) 
and the transversality of the spherical vector harmonics.
\be
\delta G_{ab} = 0 \, . 
\label{del-Gab}
\ee
The ODG expressions (\ref{del-GAb})-(\ref{del-Gab})
together with Eqs. (\ref{GE-6a}) and (\ref{GE-6b})
evaluated in the ODG yield the desired formulae 
(\ref{GE-ODG}).

\section{Stationary solutions of RW-type equations}
\label{App-B}

We discuss the conditions under which
the stationary RW type differential equation,
\be
\left[ -\p_r N(r) \p_r + \frac{1}{r^2} V(r) \right] \Psi = 0 ,
\label{RWTYPE}
\ee 
admits only the trivial solution. The potential $V(r)$ and the
function $N(r)$ are assumed to be non-negative for $r_H<r<\infty$, 
and to have analytical expansions of the form
\bdm
N(r) = N_1(r-r_H) + {\cal O}(r-r_H)^2,\;\; 
V(r) = V_H + {\cal O}(r-r_H)
\edm
in the vicinity of the horizon $r_H$, and
\bdm
N(r) = 1 + {\cal O}(r^{-1}), \;\; 
V(r) = \ell(\ell + 1) + {\cal O}(r^{-1})
\edm
as $r \rightarrow \infty$, where $\ell > 0$,
$N_1 \neq 0$, $V \neq 0$.
[In particular, the RW equation (\ref{US-5}) meets the above conditions, 
and so does the Zerilli equation, describing vacuum 
perturbations with even parity.] Under the above conditions 
the differential equation (\ref{RWTYPE}) has regular 
singular points \cite{Walter} at $r = r_H$ and $r = \infty$, 
implying that $\Psi$ behaves like
\bdm
\Psi = \left\{ \begin{array}{l}
P_1(r-r_H)  \\
\log(r-r_H) \, Q_1(r-r_H)
\end{array} \right. \; \;
\mbox{for $r \rightarrow r_H$} , 
\edm
and
\bdm
\Psi = \left\{ \begin{array}{l}
r^{-\ell - 1} \, P_2(r^{-1}) \\
r^{\ell} \, Q_2(r^{-1})
\end{array} \right. \; \;
\mbox{for $r \rightarrow \infty$} . 
\edm
Here, the $P_{1,2}(x)$ and $Q_{1,2}(x)$ are locally 
convergent power series with $P_{1,2}(0)\neq 0$ and 
$Q_{1,2}(0)\neq 0$. 

For non-negative $N(r)$ and $V(r)$ the standard integral argument,
\bea
0 &\leq& \int_{r_1}^{r_2} \left( N (\p_r\Psi)^2 + 
\frac{1}{r^2} V \Psi^2 \right) \diff r \nonumber\\
  & =  & \int_{r_1}^{r_2} \left( -\p_r N \p_r\Psi + 
\frac{1}{r^2} V \Psi \right) \Psi\,\diff r + 
[ N\Psi\p_r\Psi ]_{r_1}^{r_2},\nonumber
\eea
implies that Eq. (\ref{RWTYPE}) has only the trivial solution
$\Psi = 0$, provided that the boundary term vanishes in the limit 
$r_1\rightarrow r_H$ and $r_2\rightarrow \infty$. In particular,
this is the case if the asymptotic flatness and regularity conditions
imply that the solutions with $Q_1$ and $Q_2$ must be excluded.
 
As an example, stationary solutions of the RW equation 
(\ref{US-5}) with $\ell \geq 2$ can be excluded as follows:
The variation of the curvature components 
$\delta R_{ABcd}^{inv} \diff x^c \we \diff x^d = 
2 R^2 \diff (H / R^2) \nabhat_{[B} S_{A]}$ must be bounded,
implying that $\Psi/R$ must remain bounded as well.
Hence, the solution with $Q_1$ is not admissible, and neither is the 
one with $Q_2$, unless $\ell = 1$.

\section{Einstein-Yang-Mills background solutions}
\label{App-C}
In this Appendix we recall the behavior of the static,
spherically symmetric soliton and black hole solutions 
to the EYM equations (\ref{BK1})-(\ref{BK3}) at the singular
points (see, e.g., \cite{Forgacs}):
In the vicinity of the origin one has (with $G=1$)
\bea
N(r) &=& 1 - 4\, b^2 r^2 + O(r^4), \nonumber\\
S(r) &=& S_0 \left[ 1 + 4\, b^2 r^2 + O(r^4) \right], \label{EYMEXP1}\\
w(r) &=& 1 - b\, r^2+O(r^4), \nonumber
\eea
with parameters $b=-\frac{1}{2}w''(0)$ and $S_0 > 0$.
In the asymptotic regime one finds
\bea
N(r) &=& 1 - 2M r^{-1} + O(r^{-2}), \nonumber\\
S(r) &=& 1 + O(r^{-4}), \label{EYMEXP2}\\
w(r) &=& \pm \left[1 - \gamma 2M r^{-1} + O(r^{-2}) \right], \nonumber
\eea
with parameters $M$ and $\gamma$. Finally, in the vicinity of
the horizon the behavior is given by
\bea
N(r) &=& \frac{F_H}{r_H} (r-r_H) + O(r-r_H)^2, \nonumber\\
S(r) &=& S_H\left[ 1 + \frac{G_H^2}{r_H F_H^2} (r-r_H) + 
O(r-r_H)^2 \right], 
\label{EYMEXP3}\\
w(r) &=& w_H + \frac{G_H}{F_H} (r-r_H) + O(r-r_H)^2, \nonumber
\eea
where $F_H=1 - (w_H^2-1)^2/r_H^2$ and $G_H=w_H(w_H^2-1)/r_H$. 
Here, the free parameters are $w_H \equiv w(r_H)$ and $S_H$.

\section{su(2)-valued harmonic one-forms}
\label{App-D}

We construct a basis of su(2)-valued spherical harmonic one-forms 
which transform canonically under the angular momentum operator $J$, 
defined by
\bdm
J_X \bbT \equiv {\cal L}_X \bbT ,
\edm
where $\bbT$ is a tensor field over the {\it spherically symmetric\/} 
(pseudo-)Riemannian manifold $(M,\gtens)$. 
Here ${\cal L}_{X}$ denotes the Lie derivative with respect to 
an infinitesimal rotation $X$ on $M$. 
[In particular, for infinitesimal rotations in $\RR^3$ about the 
$x^k$-axis, we define $J_k \equiv J_{X_k}$, where 
$(X_k)_{rs} \equiv \epsilon_{krs}$].

Using the commutator relations
\be
[J_X,\diff ]=0,\;\;\; [J_X,\hatast]=0,\;\;\; [J_X, \diff r]=0 ,
\label{J_COM}
\ee
where $X$ is an infinitesimal rotation, and hence a Killing field for
$\gtens$, it is not difficult to see that
\bdm
Y\diff r,\;\;\;  \diff Y,\;\;\;  -\hatast\diff Y
\edm
form a basis of spherical harmonic {\it one-forms\/} with 
{\it total\/} angular momentum $\ell$, where $Y \equiv Y^{\ell m}$
are the standard {\it scalar\/} spherical harmonics.
The dual basis, 
$\ul{C}_1\equiv Y \ul{e}_r$, 
$\ul{C}_2\equiv \ghat^{AB} \nabhat_B Y \ul{e}_A$,
$\ul{C}_3\equiv \hat{\eta}^{AB} \nabhat_B Y \ul{e}_A$
is a linear combination of the standard {\it vector\/} harmonics
(see, e.g. \cite{Varshalovich}). 
(Here and in the following, $\ul{e}_k$ denote the standard basis
fields of $\RR^3$, $\ul{e}_r$ is the radial unit vector, and 
$\ul{e}_A$ is a basis of $S^2$ with dual basis $\hat{\theta}^A$. 
The antisymmetric tensor $\hat{\eta}_{AB}$ is defined by 
$\hatast \hat{\theta}^A = \hat{\eta}^A_{\, B} \hat{\theta}^B$.)
Since the operator $\diff$ is parity preserving, while the operator 
$\hatast$ is parity reversing, $\ul{C}_1$ and $\ul{C}_2$ have 
{\it even\/} parity, while $\ul{C}_3=\ghat^{AB} S_A \ul{e}_B$ has 
{\it odd\/} parity. Here $S_A \equiv \hat{\eta}_{AB} \hat{\nabla}^B Y$
denote the transverse spherical vector harmonics,
$\hat{g}^{AB} \hat{\nabla}_B S_A = 0$.

In order to construct su(2)-valued spherical harmonics, we use the
isometry $\ul{e}_k \leftrightarrow \tau_k$ to identify $\RR^3$ with 
su(2), where the standard inner product on $\RR^3$ corresponds to 
the normalized inner product $\mbox{Tr} \equiv -2 \mbox{trace}$ on 
su(2). Vector-valued tensors are identified with su(2)-valued 
tensors, and the operator $\diff$ is defined by the 
exterior derivative ${\cal D}$ for vector-valued forms 
$\ul{\alpha} = \alpha^i \ul{v}_i$: 
$\diff \ul{\alpha} = \ul{v}_i {\cal D}\alpha^i$ $=$
$\ul{v}_i (\diff \alpha^i + \omega^i_{\, j}\wedge\alpha^j)$, where
$\omega^i_{\, j}$ is the Riemannian connection with respect to the 
standard metric on $\RR^3$. (With respect to the standard basis, 
$\ul{v}_i = \ul{e}_i$, one has $\omega^i_{\, j} = 0$, and thus 
${\cal D} \ul{\alpha} = \diff \ul{\alpha}$, whereas, with respect to 
the basis vectors $\ul{e}_r$ and $\ul{e}_A$, one finds 
$\omega^A_{\, r} = \hat{\theta}^A$, 
$\omega^r_{\, A} = -\ghat_{AB}\hat{\theta}^B$, and 
$\omega^A_{\, B} = \hat{\omega}^A_{\, B}$.) 
The basis of su(2)-valued spherical harmonics becomes
\be
X_1 = Y\,\taur , \;\;\; 
X_2 = \ghat^{AB} \tau_A \nabhat_B Y , \;\;\; 
X_3 = \hat{\eta}^{AB} \tau_A \nabhat_B Y,
\label{SU2VSH}
\ee
that is, $X_j \equiv \ul{C}_j \cdot \ul{\tau}$. Since
the parity operator does not act on the inner index, 
$X_1$ and $X_2$ have {\it odd\/} parity, while $X_3$ has {\it even} parity.

A basis of su(2)-valued spherical harmonic {\it one-forms\/} 
is now obtained by the same procedure as above: 
Using the commutator relations 
(\ref{J_COM}), with $\diff$ generalized as above, one obtains
the nine basis vectors $\diff X_k$, $\hatast\diff X_k$,
$X_k \diff r$. The decomposition 
$\diff\alpha = \tau_i {\cal D}\alpha^i = 
\nabhat\alpha - \taur\ghat_{AB}\hat{\theta}^B \wedge \alpha^A$ 
of the total exterior derivative of a vector valued form 
$\alpha$ tangential to $S^2$ now yields the identities
\bea
\diff X_1 & = & Y \diff \taur + \taur\diff Y \, , \nonumber\\
\diff X_2 & = & \nabhat X_2 - \taur\diff Y \, , \label{XID-1}\\
\diff X_3 & = & \nabhat X_3 + \taur\hatast\diff Y \, . \nonumber
\eea
Furthermore, one has
\be
-\hatast\nabhat X_3 + \nabhat X_2 + \ell(\ell + 1) Y\diff\taur = 0.
\label{XID-2}
\ee
By virtue of these identities one may also use the one-forms 
$Y \diff \taur$, $\taur \diff Y$, $\nabhat X_2$ instead of
$\diff X_1$, $\diff X_2$, $\hatast \diff X_3$, or the one-forms 
$Y \hatast\diff \taur$, $\taur \hatast\diff Y$, $\hatast\nabhat X_2$ 
instead of $\hatast\diff X_1$, $\hatast\diff X_2$, $\diff X_3$.
In fact, the new sets turn out to be more convenient in order 
to derive the perturbation equations.

In conclusion, the su(2)-valued spherical harmonic basis one-forms
with odd parity are
\be
X_1 \diff r,\;\;\;  X_2 \diff r,\;\;\;
Y \diff \taur,\;\;\;  \taur \diff Y,\;\;\;  \nabhat X_2 ,
\label{SU2H}
\ee
while the even parity basis one-forms are
\be
X_3 \diff r,\;\;\;
Y \hatast\diff \taur,\;\;\;  \taur \hatast\diff Y,\;\;\;  \hatast\nabhat X_2 .
\label{SU2Hbis}
\ee
This is, however, only true for $\ell > 1$. For $\ell = 1$ and 
$\ell = 0$ the above fields are not linearly independent.
For $\ell = 1$ the dimensions of both the odd and the even parity sectors 
are reduced by one, since $\nabhat_{A} \nabhat_{B} Y^{(\ell = 1)}$ $=$ 
$-\ghat_{AB}Y^{(\ell = 1)}$ implies 
$\nabhat X_{2} = \ghat^{BC} \nabhat_{A} \nabhat_{B} Y \tau_{C} 
\hat{\theta}^{A} = - Y \tau_{A}\hat{\theta}^{A} = - Y \diff \taur$.
For $\ell=0$, $Y$ is constant, and hence 
$X_2$, $X_3$, and $\diff Y$ vanish. Specially, in the even parity case
only $\hatast \diff \taur$ survives, which yields the spherically
symmetric magnetic Witten ansatz for the gauge potential.

It is also worthwhile noticing that the odd-parity expansion 
(\ref{GE-2}) of the metric perturbations can be
obtained by ``lowering the inner index'' and symmetrizing the 
one-forms (\ref{SU2Hbis}):
\bea
&&X_3 \diff r = \ghat^{AB} S_A \tau_B \diff r 
\rightarrow
\delta\gfourtens = S_A (\diff r \otimes \hat{\theta}^{A} + 
\hat{\theta}^A \otimes \diff r), \nonumber\\
&&Y \hatast\diff \taur = Y \tau_A \hat{\eta}^A_{\, B} \hat{\theta}^B
\rightarrow \delta\gfourtens = 0, \nonumber\\
&&\taur \hatast\diff Y = \taur S_A \hat{\theta}^A
\rightarrow
\delta\gfourtens = S_A (\diff r \otimes \hat{\theta}^{A} + 
\hat{\theta}^A \otimes \diff r), \nonumber\\
&&\hatast\nabhat X_2 = \ghat^{BC} \tau_B \nabhat_C S_A \hat{\theta}^A
\rightarrow
\delta\gfourtens = \nabhat_{\{A} S_{B\}} 
\hat{\theta}^A \otimes \hat{\theta}^B.
\nonumber
\eea
In a similar manner the even-parity metric expansion can be obtained 
from the (odd-parity) one-forms (\ref{SU2H}).

\section{Invariant Yang-Mills perturbations}
\label{App-E}

In this Appendix we construct the gauge- and co\-or\-di\-nate-in\-va\-ri\-ant
amplitudes parameterizing the perturbations of the YM 
potential $\delta A$. Starting with Eqs. (\ref{YMP-3}), (\ref{YMP-4}) 
and (\ref{YMP-4bis}), our aim is to show that the physical perturbations
for $\ell > 1$, $\ell = 1$ and $\ell = 0$ are given by the expressions
(\ref{invpertset2}), (\ref{invpertset1}) and 
(\ref{l=0gauge}), respectively.

Under YM gauge transformations one has
\bdm
\delta A \rightarrow \delta A + \Diff \chi \, ,
\edm
where $\Diff$ is the gauge covariant derivative with respect to the 
background connection (\ref{YMBG-1}), and $\chi$ denotes the
su(2)-valued scalar field parameterizing the gauge freedom. For odd 
parity perturbations $\chi$ is given in terms of two functions
on $\tilde{M}$, 
\bdm
\chi = f_{1} X_{1} + f_{2} X_{2} ,
\edm
where $X_1$ and $X_2$ are the odd-parity scalar isospin harmonics
defined in Eq. (\ref{SU2VSH}).

Now using the identities (\ref{XID-1}) and (\ref{XID-2}) one
finds $\Diff X_1 = \taur \diff Y + w Y \diff \taur$,
$\Diff X_2 = \hat{\nabla} X_2 - w \taur \diff Y$,
the amplitudes defined in Eqs. (\ref{YMP-3}) and (\ref{YMP-4}) are
found to behave as follows under gauge transformations:
\be
\left. \begin{array}{l}
\alpha \rightarrow \alpha + \diff f_{1} \\
\beta  \rightarrow  \beta + \diff f_{2} \\
\mu    \rightarrow    \mu + f_{1} -  f_{2}w
\end{array} \right. 
\; \; \mbox{for $\ell \geq 1$,}
\label{YMP-7a}
\ee
and
\be
\nu \rightarrow \nu + f_{1}w -f_{2}
\; \; \; \mbox{for $\ell = 1$,}
\label{YMP-7b}
\ee
\be
\left. \begin{array}{l}
\nu \rightarrow \nu + f_{1}w \\
\sigma \rightarrow \sigma + f_{2}
\end{array} \right. 
\; \; \mbox{for $\ell > 1$.} 
\label{YMP-7c}
\ee

For $\ell = 1$, one can introduce two gauge-invariant one-forms 
$a$ and $b$, say,
\be
\left. \begin{array}{l}
a  \equiv  \alpha - \diff 
\left(\frac{\mu - w \nu}{1-w^{2}} \right) \\
b  \equiv  \beta + \diff 
\left(\frac{\nu - w \mu}{1-w^{2}} \right) 
\end{array} \right.  
\; \; \mbox{for $\ell = 1$,}
\label{YMP-8}
\ee
which are well-defined unless the background configuration is the
Schwarzschild black hole, 
$w = 1$. The transformation laws (\ref{YMP-7a}) and (\ref{YMP-7b}) 
imply that there exists a gauge for which the scalars $\mu$ and 
$\nu$ vanish. Moreover, the above definitions show that in this 
gauge the one-forms $\alpha$ and $\beta$ coincide with the gauge 
invariant one-forms $a$ and $b$. Since the perturbation equations
are gauge-invariant, we may thus parametrize $\delta A^{(\ell = 1)}$ 
in terms of the two gauge-invariant one-forms $a$ and $b$ on 
$\tilde{M}$,
\be
\delta A^{(\ell = 1)} = X_{1} a + X_{2} b .
\label{YMP-4-GI}
\ee

For $\ell > 1$, we may proceed in a similar way and introduce two 
gauge-invariant one-forms and one gauge-invariant function as follows:
\be
\left. \begin{array}{l}
a  \equiv  \alpha - \diff 
\left( \mu + w \sigma \right) \\
b  \equiv  \beta - \diff \sigma \\
c   \equiv  \nu - w \left( \mu + w \sigma \right)
\end{array} \right.   
\; \; \mbox{for $\ell > 1$.}
\label{YMP-9}
\ee
It is again obvious from Eqs. (\ref{YMP-7a}) and (\ref{YMP-7c}) 
that there exists a gauge for which $\mu$ and $\sigma$ vanish,
and that the remaining amplitudes $\alpha$, $\beta$ and $\nu$
coincide with the  gauge-invariant quantities $a$, $b$ and $c$ 
in this gauge. Hence, without loss of generality, we may set
\be
\delta A^{(\ell > 1)} = X_{1} a + X_{2} b + c \,
Y \diff \taur ,
\label{YMP-3-GI}
\ee
and consider $a$, $b$ and $c$ as gauge-invariant amplitudes.

For $\ell = 0$, $\delta A$ is parametrized in terms of the 
one-form $\alpha$ and the function $\nu$, which transform 
according to $\alpha \rightarrow \alpha + \diff f_{1}$ and
$\nu \rightarrow  \nu + f_{1}w$, respectively. The amplitudes 
combine into a gauge-invariant one-form
\be
a \equiv \alpha - \diff \left(\frac{ \nu}{w} \right)
\; \; \; \mbox{for $\ell = 0$,}
\label{1-ell-0}
\ee
where $\alpha$ coincides with $a$ in the gauge for which $\nu$ 
vanishes. (This gauge does not exist for the RN 
background, since $\nu$ is gauge-invariant for $w(r) \equiv 0$.) 
In terms of $a$ one has
\be
\delta A^{(\ell = 0)} =  \taur \, a .
\label{2-ell-0}
\ee

So far we have parametrized $\delta A$ in terms of 
gauge-invariant amplitudes, or, more precisely, in terms of 
amplitudes which coincide with gauge-invariant amplitudes 
in a certain gauge. However, these quantities 
are not yet invariant under infinitesimal coordinate 
transformations on the background. As the linearized Einstein 
and YM equations are invariant under these transformations, 
they will involve only coordinate-invariant combinations of the 
above amplitudes. In order to find these combinations, it remains 
to study the behavior of the gauge-invariant amplitudes
$a$, $b$ and $c$ under the transformation
\bdm
\delta A \rightarrow \delta A + {\cal L}_{X} A \, ,
\edm
where $A$ is the background connection given in 
Eq. (\ref{YMBG-1}), and
${\cal L}_{X}$ denotes the Lie derivative with respect to the 
infinitesimal vector field 
$X^{\mu} = - f R^{-2} \delta^{\mu}_{\; A} 
\hat{\eta}^{AB} \hat{\nabla}_{B} Y$, defined in
Eq. (\ref{GE-3}). In terms of the coordinate freedom $f$, 
one finds
\bdm
{\cal L}_{X} A = (1-w)\left[
\frac{f}{R^{2}} \left( \taur \diff Y + \nabhat X_{2}
\right) + X_{2} \, \diff \left( \frac{f}{R^{2}} \right) \right] .
\edm
(The most efficient way to establish this is to write
${\cal L}_{X} = \diff i_{X} + i_{X} \diff$, and to
use $i_{X} \diff \Omega = -R^{-2} f \diff Y$ and
$i_{X} \hat{\ast} \diff \taur = R^{-2} f X_{2}$.)

The transformation properties of the one-forms
$\alpha$, $\beta$ and the functions $\mu$, $\nu$ and $\sigma$ 
defined in Eq. (\ref{YMP-3}) are now immediately obtained. 
[For $\ell = 1$ one has to replace
$ \nabhat X_{2}$ by $-Y \diff \taur$ and to use Eq. 
(\ref{YMP-4}) instead of Eq. (\ref{YMP-3}).]
For $\ell > 1$, the gauge-invariant quantities (\ref{YMP-9})
transform as follows under coordinate transformations
generated by $X$:
\be
\left. \begin{array}{l}
a \rightarrow  a - \diff 
\left[ \frac{f}{R^{2}} \left( 1-w^{2} \right) \right] \\
b \rightarrow  b + \frac{f}{R^{2}} \, \diff w \\
c \rightarrow  c - \frac{f}{R^{2}} \, w \left( 1-w^{2} \right)
\end{array} \right.   
\; \; \mbox{for $\ell > 1$,}
\label{Lie-3}
\ee
while the transformation laws for the quantities
(\ref{YMP-8}) become
\be
\left. \begin{array}{l}
a \rightarrow a - \diff \left(\frac{f}{R^{2}}\right) \\
b \rightarrow b - w \, \diff \left(\frac{f}{R^{2}}\right) 
\end{array} \right.   
\; \; \mbox{for $\ell = 1$.}
\label{Lie-4}
\ee
(There exist no allowed 
coordinate transformations in the odd-parity sector if $\ell = 0$.) 
For $\ell > 1$, one may eventually use the transformation property
(\ref{GE-4}) of the metric perturbation $\kappa$, 
$\kappa  \rightarrow  \kappa + f$, to introduce the following 
gauge {\it and\/} coordinate-invariant amplitudes:
\bea
A & \equiv& a + \diff 
\left[ \frac{\kappa}{R^{2}} \left( 1-w^{2} \right) \right] ,
\; \; \; 
B \equiv b - \frac{\kappa}{R^{2}} \, \diff w 
\nonumber\\
C &\equiv& c + \frac{\kappa}{R^{2}} \, w \left( 1-w^{2} \right),
\; \; \; 
H  \equiv  h - R^{2} \diff \left( 
\frac{\kappa}{R^{2}} \right),
\label{Lie-5}
\eea
where we have also recalled the definition (\ref{GE-7}) of the
coordinate-invariant metric perturbation one-form $H$.
In the ODG ($\kappa = 0$) these gauge-and 
coordinate-invariant amplitudes coincide with the gauge-invariant amplitudes
$a$, $b$, $c$, and $h$, which reduce to the original amplitudes 
$\alpha$, $\beta$, $\gamma$, and $h$ in the ODSG 
($\kappa = \mu = \sigma = 0$).

For $\ell = 1$ the gauge- and coordinate-invariant YM
amplitudes are obtained by comparing the transformation laws 
(\ref{Lie-4}) with the transformation property (\ref{GE-4}) 
of the metric perturbation $h$, 
$h \rightarrow h + R^{2} \diff (R^{-2} f)$.
This yields the invariant quantities $\bar{a}$ and $\bar{b}$,
defined by
\be
\bar{a} \equiv a + \frac{h}{R^2} , \;\;\; 
\bar{b} \equiv b + w \frac{h}{R^2} .
\label{Lie-6}
\ee

\section{Linearized flux integrals}
\label{App-F}

The Komar expressions for the local electric and magnetic charges, 
the local mass and the local angular momentum of a stationary spacetime
are given by the following flux integrals over a sphere with 
radius $R$:
\bea
&&Q_e(R) = \frac{1}{4\pi} \int_{S_R}\ \ast F, \;\;\;
Q_m(R) = \frac{1}{4\pi} \int_{S_R}\ F, \nonumber\\
&&M(R) = -\frac{1}{8\pi G} \int_{S_R}\ast
\left( \diff g_{t\mu}\wedge\diff x^{\mu} \right), \nonumber\\
&&J(R) = \frac{1}{16\pi G} \int_{S_R}\ast
\left( \diff g_{\varphi\mu}\wedge\diff x^{\mu} \right). \nonumber
\eea
Using the expressions (\ref{invpertset2}) and (\ref{invpertset1})
for the gravitational and the YM perturbations, 
the linearized flux integrals are found to be
\bea
&&\delta Q_m(R) = \delta M(R) = 0, \nonumber\\
&&\delta Q_e(R) \sim   
\delta_{\ell \,1} \ul{e}_m\cdot\ul{\tau} \,
R^2\left( \tilast F_A + 2\tilast F_B \right), \nonumber\\
&&\delta J(R) \sim \delta_{\ell \,1} 
\delta_{m \,0} R^4\tilast F_H,
\label{FLUX}
\eea
where $\ul{e}_0 = (0,0,1)$, $\ul{e}_\pm = (\mp 1,i,0)$, 
and $F_H$, $F_A$ and $F_B$ are defined in Eq. (\ref{twoforms}).
Here we have also used the orthogonality of 
the spherical harmonics $Y^{\ell m}$ and the expansions 
$\ul{e}_r \sim Y^{1m} \ul{e}_m$ and 
$S_\varphi^{\ell m} \sim \ghattens(\diff Y^{\ell m}, \diff Y^{10})$.

\end{document}